\documentclass[reprint,aps,pra,twocolumn,superscriptaddress]{revtex4-2}
\usepackage[T1]{fontenc}
\usepackage{amsmath, amsthm, amssymb,amsfonts,mathbbol,amstext}
\usepackage{bm}
\usepackage{bbm}
\usepackage{bbold}
\usepackage{comment}
\usepackage{color}
\usepackage{diagbox}
\usepackage{dcolumn}
\usepackage{float}
\usepackage{graphicx}
\usepackage{hyperref}
\usepackage{mathtools}
\usepackage{multirow}
\usepackage{nccmath}
\usepackage{pifont}
\usepackage{soul}
\usepackage{xcolor}  
\usepackage{fancyhdr}
\usepackage{eso-pic}
                    
\graphicspath{./Figure}
\newcommand{\bra}[1]{\langle#1|}
\newcommand{\ket}[1]{|#1\rangle}

\newcommand{\proj}[1]{\ket{#1}\!\bra{#1}}
\newcommand{\tr}{{\rm tr}}
\newcommand{\mc}{\mathcal}

\begin{document}
    \preprint{APS/123-QED}

 \title{Linear-optical test of quantum contextuality with sequential measurements} 

\author{Jiaqi Liu}
    \thanks{These authors contributed equally to this work as co-first authors.}
    \affiliation{National Laboratory of Solid State Microstructures, Key Laboratory of Intelligent Optical Sensing and Manipulation, College of Engineering and Applied Sciences and School of Physics, Collaborative Innovation Center of Advanced Microstructures, Nanjing University, Nanjing, 210093, China}
    
\author{Bita Olamaei}
    \thanks{These authors contributed equally to this work as co-first authors.}
    \affiliation{Physics Department, Institute for Advanced Studies in Basic Sciences (IASBS), Gava Zang, P.O. Box 45137-66731, Zanjan, Iran}    
    \affiliation{Vienna Center for Quantum Science and Technology, Atominstitut, TU Wien, 1020 Vienna, Austria}

\author{Lijian Zhang}
    \email{lijian.zhang@nju.edu.cn}
    \affiliation{National Laboratory of Solid State Microstructures, Key Laboratory of Intelligent Optical Sensing and Manipulation, College of Engineering and Applied Sciences and School of Physics, Collaborative Innovation Center of Advanced Microstructures, Nanjing University, Nanjing, 210093, China}
    
\author{Ali Asadian}
    \email{ali.asadian@iasbs.ac.ir}
    \affiliation{Physics Department, Institute for Advanced Studies in Basic Sciences (IASBS), Gava Zang, P.O. Box 45137-66731, Zanjan, Iran}
    
\author{Saleh Rahimi-Keshari}
    \email{keshari@ipm.ir}
    \affiliation{School of Quantum Physics and Matter, Institute for Research in Fundamental Sciences (IPM), P.O. Box 19395-5531, Tehran, Iran}    
    \affiliation{Wallenberg Centre for Quantum Technology, Department of Microtechnology and Nanoscience, Chalmers University of Technology, SE-412 96 Gothenburg, Sweden}

    \date{\today}

\begin{abstract}
     Quantum contextuality provides a fundamental signature of nonclassical behavior that cannot be explained by noncontextual hidden-variable models. We propose and experimentally implement a linear-optical setup for demonstrating Kochen--Specker contextuality via a violation of the KCBS inequality using single photons. Our scheme employs sequential measurements realized with linear-optical networks and on--off photodetectors. The construction ensures that each co-measured observable is implemented by the same physical operation across different contexts. Our experimental results demonstrate a clear violation of the KCBS inequality and robustness against photon loss. Beyond fundamental investigations, the proposed setup provides a practical tool for probing non-classicality and photon-number statistics of quantum states, which in turn enables the verification of single-photon sources.       
\end{abstract}

\maketitle

\section{Introduction} \label{Sec:Intro}

The dependence of measurement outcomes on the measurement context—the set of jointly measured compatible observables—is known as contextuality \cite{kochen1967the, bell1966problem, budroni2022kochen}. This phenomenon, originally identified in the works of Bell, Kochen, and Specker \cite{bell1966problem, kochen1967the, Specker1960-SPEDLN}, reveals a fundamental departure between the predictions of quantum theory and those of classical descriptions based on noncontextual hidden-variable models. A prominent example is provided by Bell tests, which illustrate a particular form of contextuality known as Bell nonlocality \cite{Nonlocality}. In recent years, contextuality has also been recognized as an important resource for a variety of quantum information processing tasks. These include applications in quantum communication \cite{nagata2005kochen, saha2019preparation, gupta2023quantum}, quantum state discrimination \cite{PRXschmid}, quantum computation \cite{howard2014contextuality, delfosse2015wigner, dawkins2015qutrit, raussendorf2013contextuality, bermejo2017contextuality, raussendorf2017contextuality}, and the device-independent certification of system properties such as dimensionality \cite{dimensionWitness, SelfTesting}.

In the Kochen--Specker framework \cite{kochen1967the}, the physical assumptions underlying noncontextual hidden-variable (classical) models can be translated into experimentally testable statistical constraints in the form of inequalities, known as noncontextuality inequalities, which bound the correlations that can be observed in a classical theory. Experimental evidence of contextuality is obtained through violations of these inequalities. A variety of such inequalities have been proposed \cite{budroni2022kochen, klyachko2008simple, yu2012state, clauser1969proposed, mermin1990simple, mermin1993hidden, peres1990incompatible, peres1991two}, applicable to both single systems and multipartite settings. These inequalities are satisfied by noncontextual classical models, while they can be violated by quantum mechanics.

One of the simplest examples is the inequality introduced by Klyachko, Can, Binicio{\u{g}}lu, and Shumovsky (KCBS) \cite{klyachko2008simple}, which involves correlation measurements performed on a single physical system. The inequality is built from five measurements $\{A^{(j)}\}$, $j=1,...,5$, with correlations taken only between neighboring compatible pairs $A^{(j)}$ and $A^{(j+1)\hspace{0.3em} mod\hspace{0.3em} 5}$. Since its proposal, the KCBS inequality has attracted significant theoretical and experimental attention. The first experimental test was carried out using photons~\cite{lapkiewicz2011experimental}. Subsequent demonstrations have been realized in a variety of physical platforms, including trapped ions \cite{kirchmair2009state}, neutrons \cite{bartosik2009experimental}, and superconducting qubits \cite{jerger2016contextuality}.  Among these platforms, photonic systems offer several advantages, including the ability to realize high-dimensional quantum states with high fidelity and low decoherence, as well as well-established optical control techniques at relatively low cost.

However, testing contextuality with photons remains challenging due to the difficulty of operationally defining measurement contexts and the destructive nature of photodetection, which poses a major obstacle to sequential measurements required for the accurate evaluation of joint observables. One way to circumvent this issue is to avoid a physical detector for the first measurement and instead encode its outcome into different degrees of freedom, such as polarization, paths \cite{huang2003experimental, SICphoton, ahrens2013two, d2013experimental, amselem2012experimental, arias2015testing,zu2012state,borges2014quantum}, or temporal modes \cite{ahrens2013two}. In this strategy, the measurement outcomes are effectively combined and read out collectively at the end of the experiment, thereby realizing a joint measurement scheme rather than a sequence of measurements. This approach, however, does not fully capture the quantum nature of contextuality, as noted in \cite{asadian2022bosonic}, and yields classically simulable data \cite{Clcorr}. Other experiments \cite{lapkiewicz2011experimental, zhang2019experimental} address this issue by evaluating joint measurements of two observables through correlations between corresponding pairs of detectors. Nevertheless, these implementations do not ensure that the same observable is realized identically across different contexts. In particular, instead of measuring the compatible observables $A^{(5)}A^{(1)}$, the experiment effectively measures $A^{(5)}A'^{(1)}$, where the observable $A'^{(1)}$ is physically distinct from the observable $A^{(1)}$ appearing in the correlation $A^{(1)}A^{(2)}$.

In this paper, we propose a simple linear-optical scheme for testing quantum contextuality using linear-optical networks and on--off photodetectors. Similar to Refs.~\cite{lapkiewicz2011experimental, zhang2019experimental}, our implementation employs two distinct photodetectors associated with the two measurement outcomes. However, our construction leverages no-click events corresponding to vacuum detection to realize a sequential measurement scheme rather than a joint one, hence fulfilling the key requirements of repeatability and non-disturbance in contextuality tests. Crucially, our approach ensures that each observable is implemented by the same physical operation whenever it appears in different measurement contexts, hence resolving the long-standing issue of realizing identical observables across contexts. In addition, our scheme can be used to verify nonclassicality and probe the photon-number components of quantum states. Using this construction, we experimentally demonstrate a violation of the KCBS inequality using single photons. Moreover, we observe the violation with up to $10.55\%$ loss, which confirm the robustness of our approach and establishing the feasibility.

This paper is structured as follows. Section \ref{Sec:KCBS} reviews the KCBS noncontextuality inequality and its simplified forms. Section \ref{Sec:Methods} introduces our linear-optical setup. In Section \ref{Sec:Verification}, we discuss applications of our setup for verification of nonclassicality and photon-number statistics of quanutm states. Section \ref{Sec:Experiment} details the experimental apparatus, and Section \ref{Sec: Results} presents the results, which demonstrate a clear violation of the KCBS inequality. Finally, Section \ref{Sec:Conclu} provides conclusions and an outlook.

\section{The KCBS Inequality} \label{Sec:KCBS}

We consider a quantum system with a Hilbert space of dimension at least three and $N$ distinct measurement settings. Each setting $j \in \{1, \dots, N\}$ corresponds to a projective measurement with two possible outcomes, described by the projection operators $\Pi^{(j)}_{-}$ and $\Pi^{(j)}_{+}.$ Here, the superscript $(j)$ labels the measurement setting, while the subscripts indicate the corresponding outcome. These operators satisfy the completeness relation
$\Pi^{(j)}_{-}+\Pi^{(j)}_{+}=\mathbb{1}$, where $\mathbb{1}$ denotes the identity operator.

To describe the measurement statistics, we introduce a random variable $v_j \in \{-1, +1\}$ representing the outcome of measurement $j$. According to the Born rule, the probability of obtaining outcome $v_j$ for the measurement setting $j$ is
\begin{equation}
    p(v_j) = \mathrm{Tr}\!\left[ \rho \,\Pi^{(j)}_{v_j} \right],
\end{equation}
where $\rho$ is the quantum state of the system. Using these projection operators, we define a corresponding set of observables
\begin{equation}
    A^{(j)} = \sum_{v_j \in \{-1,+1\}}\!\!\!\!\! v_j \Pi^{(j)}_{v_j}
    = \Pi^{(j)}_{+} - \Pi^{(j)}_{-},
\end{equation}
which are used to construct the contextuality tests.

One of the simplest state-dependent noncontextuality inequalities was proposed by Klyachko, Can, Binicio\u{g}lu, and Shumovsky \cite{klyachko2008simple}. The KCBS inequality is constructed from five dichotomic measurements $\{A^{(j)}\}_{j=1}^{5}$ with outcomes $\{-1,+1\}$, arranged such that only adjacent measurements are compatible. Specifically, the commutation relations $[A^{(j)},A^{(j-1)}]=[A^{(j)},A^{(j+1)}]=0$, hold, with cyclic notation implied. Each pair of neighboring measurements therefore defines a measurement context.

According to the Kochen--Specker theorem, if measurement outcomes reveal predetermined values, then those outcomes must be independent of the measurement context, namely, of whether $A^{(j)}$ is jointly measured with $A^{(j-1)}$ or with $A^{(j+1)}$. However, when measurement outcomes are not predetermined, this assumption fails, leading to distinguishable predictions between quantum mechanics and noncontextual classical models. The KCBS inequality directly tests this assumption of noncontextuality. For any noncontextual hidden-variable model, the inequality
\begin{equation}\label{eq.KCBS}
    \kappa(\rho) = \sum_{j=1}^5 \big\langle A^{(j)} A^{(j+1)} \big\rangle \geq -3,
\end{equation}
where $\big\langle A^{(j)} A^{(j+1)} \big\rangle=\tr\big[\rho A^{(j)} A^{(j+1)} \big]$, must hold \cite{klyachko2008simple}. The classical lower bound of $-3$ arises because all possible assignments of predetermined values $\pm 1$ to the five measurements satisfy the inequality algebraically.

Crucially, noncontextual hidden-variable models assume the existence of a global joint probability distribution $p(v_1,\ldots,v_5)$ from which all observable joint probabilities $p(v_i,v_j)$ are obtained as marginals, as formalized by Fine’s theorem~\cite{FinePRL1982}. Experimentally, testing the KCBS inequality therefore requires nondisturbing measurements, such that the outcome statistics of any single measurement can be recovered from compatible sequential measurements. This requirement is operationally captured by the no-signaling-in-time condition, which ensures that the outcome statistics of any single measurement are independent of whether a compatible measurement was performed earlier, thereby guaranteeing the consistency of marginal probabilities across different sequential measurement contexts.

To satisfy the compatibility requirement between neighboring measurements, one can define the projection operators
$\Pi^{(j)}_- = \proj{\psi^{(j)}}$ corresponding to the outcome $-1$, where the states $\{\ket{\psi^{(j)}}\}_{j=1}^5$ are chosen such that the adjacent states are orthogonal. With this choice, the set of two-outcome projective measurements can be written as
\begin{equation} \label{Eq:Obs.A^{(j)}}
    A^{(j)} = \Pi^{(j)}_+ - \Pi^{(j)}_- = \mathbb{1} - 2 \proj{\psi^{(j)}},
\end{equation}
where $\Pi^{(j)}_+ = \mathbb{1} - \proj{\psi^{(j)}}$ corresponds to the $+1$ outcome.

According to the Kochen--Specker theorem, the simplest quantum system capable of demonstrating contextuality is a qutrit, whose Hilbert space has dimension three. For a qutrit, using a spherical parametrization of the coefficients, the state vectors can be expressed as $\ket{\psi^{(j)}} = \cos\theta_j \ket{r_1} + \sin\theta_j \cos\phi_j \ket{r_2} + \sin\theta_j \sin\phi_j \ket{r_3}$, where $\{\ket{r_1}, \ket{r_2}, \ket{r_3}\}$ is an orthonormal basis. The inner product between two such states is given by $\bra{\psi^{(i)}}\psi^{(j)}\rangle = \cos\theta_i \cos\theta_j + \sin\theta_i\sin\theta_j\cos(\phi_i - \phi_j)$. Following the KCBS construction, the coefficients can be chosen such that $\bra{\psi^{(j)}} \psi^{(j+1)}\rangle = 0$, which ensures the compatibility condition $[A^{(j)}, A^{(j+1)}] = 0$. This condition can be realized by choosing $ \theta_j = \cos^{-1}\!\left(\frac{1}{\sqrt[4]{5}}\right)$ and $\phi_j = \frac{4\pi j}{5}$. Note that this orthogonality also implies that the corresponding $-1$ outcomes never occur simultaneously in the correlation measurements of $A^{(j)}$ and $A^{(j+1)}$, a property known as the exclusivity condition.  
 
With these parametrizations, one can see that the maximum violation is obtained for the initial state $\ket{\psi_{\rm in}}=\ket{r_1}$ satisfying $\langle\psi^{(j)}|\psi_{\rm in}\rangle = 1/{\sqrt[4]{5}}$ for all $j$. In this case, the two-point correlations are evaluated as $\langle\psi_{\rm in} | A^{(j)} A^{(j+1)} | \psi_{\rm in}\rangle = 1 - \frac{4}{\sqrt{5}}$, leading to the KCBS sum
\begin{equation} \label{eq:max-kappa}
    \kappa(\rho) = \sum_{j=1}^5 \big\langle A^{(j)} A^{(j+1)}\big\rangle = 5 - 4 \sqrt{5} \approx -3.944.
\end{equation}
 
From an implementation perspective, it is convenient to work with an alternative formulation of the KCBS inequality expressed in terms of joint probabilities \cite{cabello2016simple},
\begin{equation}\label{eq:joint-prob}
    p(v_j, v_{j+1})=\tr \big[\Pi^{(j+1)}_{v_{j+1}} \Pi^{(j)}_{v_j} \rho \Pi^{(j)}_{v_j} \Pi^{(j+1)}_{v_{j+1}} \big],
\end{equation}
where $\rho$ denotes the initial state of the system. An explicit evaluation of these joint probabilities for sequential projective measurements yields
\begin{equation}\label{Eq:Joint.Prob}
        p(v_j, v_{j+1}) = p(v_j)\,
        \tr \big[ \Pi^{(j+1)}_{v_{j+1}} \rho_{v_j} \Pi^{(j+1)}_{v_{j+1}} \big],
\end{equation}
where $\rho_{v_j} = \Pi^{(j)}_{v_j} \rho \Pi^{(j)}_{v_j}/p(v_j)$ is the post-measurement state conditioned on obtaining the outcome $v_j$ with probability $p(v_j)=\tr[\Pi^{(j)}_{v_j} \rho \Pi^{(j)}_{v_j}]$ after measuring $A^{(j)}$.

An equivalent and particularly simple reformulation of the KCBS inequality using these probabilities \cite{cabello2016simple} can then be written as
\begin{equation}\label{Eq:Simp.KCBS}
    \mathcal{S}(\rho)  = \sum_{j=1}^{5} p(v_j=+1) - \sum_{j=1}^{5} p(v_j=+1,v_{j+1}=+1) \leq 2 .
\end{equation}
The original KCBS inequality in Eq.~\eqref{eq.KCBS} can be recovered by noting that $\Pi^{(j)}_{+} =(\mathbb{1}+ A^{(j)})/2$,
from which one finds $\mathcal S(\rho) = \frac{1}{4}(5-\kappa(\rho))$; see Appendix~\ref{appen.KCBS}. Note that from Eq.~\eqref{eq:max-kappa}, the maximal quantum violation corresponds to $\mathcal{S}(\rho)=\sqrt{5}\approx 2.236$. 

In photonic systems, implementing ideal sequential measurements is particularly challenging, since photon detection is typically destructive. Although non-demolition measurement schemes exist \cite{kok2002single}, they are either probabilistic or rely on strong optical nonlinearities, making the preservation of the post-measurement state experimentally demanding and resource intensive. We overcome this difficulty by introducing tailored joint measurements together with specific outcome assignments for on--off detectors. This approach renders the simplified KCBS inequality in Eq.~\eqref{Eq:Simp.KCBS} experimentally accessible, providing a practical route to implementing sequential projective measurements on photons using only linear optics.

\section{A NEW SEQUENTIAL SCHEME FOR TESTING CONTEXTUALITY}\label{Sec:Methods}

\begin{figure}
    \centering
    \includegraphics[width=3.3in]{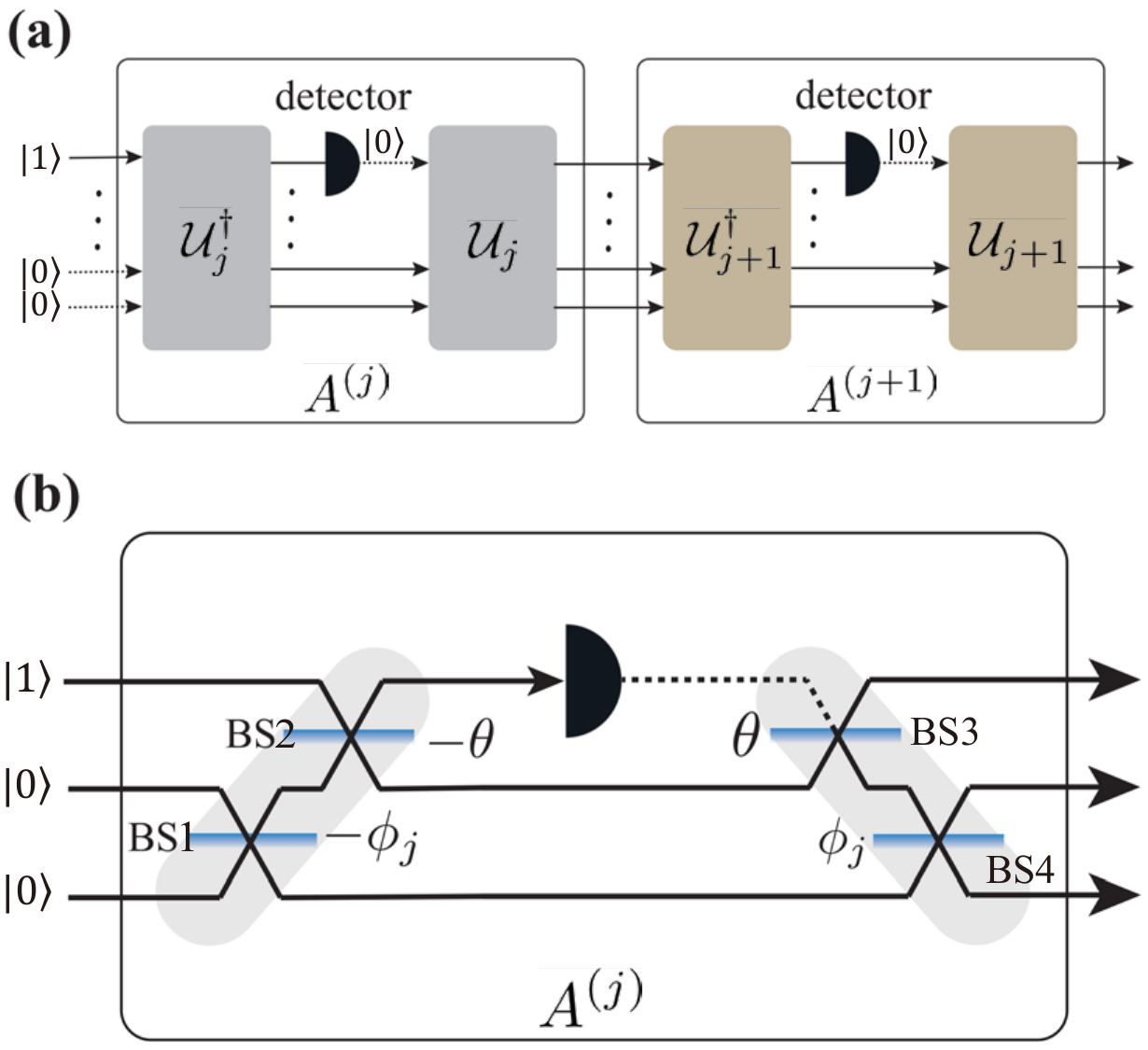}
    \caption{(a) Setup for the implementation of the KCBS noncontextuality test. Our system consists of a single photon distributed over $M$ modes, which can model any quantum system with an $M$-dimensional Hilbert space. Measurement $A^{(j)}$ is realized by an on--off photodetector sandwiched between two linear-optical networks described by unitary operators $\mathcal{U}_j$ and $\mathcal{U}_j^\dagger$. Conditioning on no-click events from the on--off photodetector projects the first mode between the networks onto the vacuum state without destroying the single photon, allowing the subsequent measurement $A^{(j+1)}$ to be performed. We show that the KCBS quantity can be evaluated by measuring the corresponding no-click probabilities. Furthermore, optimal linear-optical networks for observing the maximal violation with the initial state $\ket{1,0,\dots,0}$ can be realized using linear-optical networks with real transfer matrices ${U}^{(j)}$ whose first rows are given by Eq.~\eqref{Eq:Unitary}. This enables the implementation of the linear-optical networks using only $M-1$ beam splitters (BS). (b) Realization of the measurement $A^{(j)}$ for $M = 3$ using beam splitters with transmissivities $\cos\theta$ and $\cos\phi_j$.}
    \label{Fig:OpticalSetup}
\end{figure}

We begin by introducing the physical system and the basic elements of the experimental setup used to implement the KCBS noncontextuality test in the form given by Eq.~\eqref{Eq:Simp.KCBS}; see Fig.~\ref{Fig:OpticalSetup}. Our quantum system consists of a single photon distributed among $M\geq3$ bosonic modes. 

A single bosonic mode is described by the annihilation and creation operators $a$ and $a^\dagger$, satisfying the canonical commutation relation $[a,a^\dagger]=1$. The corresponding Hilbert space is infinite dimensional and is spanned by the eigenstates of the number operator $a^\dagger a$, namely the Fock (number) states. For an $M$-mode system, we have $M$ pairs of annihilation and creation operators $\{a_1,a_1^\dagger,\dots,a_M,a_M^\dagger\}$, which obey the commutation relations $[a_n,a_m^\dagger]=\delta_{nm}$ and $[a_n,a_m]=[a_n^\dagger,a_m^\dagger]=0$. The total Hilbert space is given by the tensor product of the individual mode Hilbert spaces and can be decomposed as a direct sum of subspaces with fixed total photon number.

In this work, we focus on the single-photon subspace of the Hilbert space associated with the $M$-mode system, which is $M$-dimensional and spanned by the states
\begin{equation*}
    \Big\{\ket{1}_n = a_n^\dagger \ket{\bm 0}
    = \ket{0,\dotsc,0,\underbrace{1}_{\text{$n$-th mode}},0,\dotsc,0}\Big\}_{n=1}^M,
\end{equation*}
where $\ket{\bm 0}$ denotes the multimode vacuum state. The state $\ket{1}_n$ corresponds to a single photon occupying the $n$-th mode. In this encoding, an arbitrary pure state in the effective $M$-dimensional Hilbert space can be written as
\begin{equation} \label{eq:state-psi}
\ket{\psi} = \sum_{n=1}^{M} c_n \ket{1}_n,
\qquad
\sum_{n=1}^{M} |c_n|^2 = 1.
\end{equation}

To describe unitary transformations, we use lossless linear-optical networks consisting of passive elements, such as beam splitters and phase shifters. Such a network is described by a unitary operator $\mc{U}$ that linearly combines the creation operators as, 
\begin{equation}\label{Eq:LON}
    \mc{U} a^\dagger_n\, \mc{U}^{\dagger} = \sum_{m = 1}^{M} U_{nm} a^\dagger_m,
\end{equation}
where $U$ is known as the transfer matrix that is a $M\times M$ unitary matrix here. From this relation, it follows that state vectors within the single-photon subspace can be transformed into one another by suitable linear-optical transformations. For instance, any arbitrary pure state in the single-photon subspace can be generated by applying an appropriate linear-optical network to the state $\ket{1}_1$,
\begin{equation}\label{eq:psi}
    \ket{\psi} = \mc{U}\ket{1}_1 =\mc{U}a^\dagger_1\mc{U}^\dagger\ket{\bm0}= \sum_{m = 1}^M U_{1m}\ket{1}_m,
\end{equation}
where we have used Eq.~\eqref{Eq:LON} and the fact that $\mc{U}\ket{\bm0}=\ket{\bm0}$, since passive linear-optical networks do not generate photons.

The measurement settings for this system can be realized using linear optical networks combined with on--off detectors. In our scheme, only the first mode of the $M$-mode system is directly measured by an on--off detector. This implements a dichotomic measurement described effectively by the projector $\Pi_+ = \sum_{n=2}^{M} \ket{1}_n\!\bra{1}$, corresponding to the outcome $+1$ (no-click), and the projector $\Pi_- = \ket{1}_1\!\bra{1}$, corresponding to the outcome $-1$ (click). These operators define the observable $D = \Pi_+ - \Pi_-$. In principle, such a measurement could be implemented in a quantum non-demolition manner using Kerr nonlinearities \cite{imoto1985quantum,sanders1999nonclassical,ou1996complementarity,boyd1999order,lukin2000nonlinear} or linear-optical schemes \cite{kok2002single}, thereby preserving the post-measurement state  after detecting the outcome $-1$. In our setup, however, photon detection is destructive. Specifically, when the measurement is applied to the single-photon state $\ket{\psi}$ defined in Eq.~\eqref{eq:state-psi}, detecting outcome $-1$ (a click event) annihilates the photon, leaving the system in the vacuum state $\ket{\bm 0}$. In contrast, upon obtaining outcome $+1$ (no click), the photon is preserved in the remaining modes, and the post-measurement single-photon state becomes $\frac{1}{\sqrt{p_+}} \sum_{n=2}^{M} c_n \ket{1}_n$, where $p_+ = 1- |c_1|^2$ is the probability of the no-click outcome. 

We now describe the implementation of the KCBS noncontextuality test in our setting. As illustrated in Fig.~\ref{Fig:OpticalSetup}(a), the five dichotomic observables required for the test are realized by applying an appropriate linear-optical unitary transformation prior to the on--off photodetection on the first mode, followed by its Hermitian conjugate after the detection stage. In this way, the measurement basis associated with the fixed on--off detector is effectively rotated, allowing us to implement different observables. We thus define
\begin{equation}\label{Eq:Obs.A^{(j)}=UA0Ud}
    A^{(j)} = \mathcal{U}_j D \mathcal{U}_j^\dagger
    = \Pi_+^{(j)} - \Pi_-^{(j)},
\end{equation}
for $j=1,\dots,5$, where $\Pi_-^{(j)} = \mathcal{U}_j \ket{1}_1\!\bra{1} \mathcal{U}_j^\dagger
= \ket{\psi^{(j)}}\bra{\psi^{(j)}}$ and $\Pi_+^{(j)} = \mathbbm{1} - \ket{\psi^{(j)}}\bra{\psi^{(j)}}$. Here, $\mathbb{1} = \sum_{n=1}^M \ket{1}_n\!\bra{1}$ denotes the identity operator in the single-photon subspace. As follows from Eq.~\eqref{eq:psi}, the state $\ket{\psi^{(j)}}$ depends only on the first row of the corresponding transfer matrix. Therefore, by appropriately choosing the elements $\{U_{1n}^{(j)}\}$ of the first row only, one can design optimal measurements satisfying the conditions discussed in Sec.~\ref{Sec:KCBS}. In particular, the condition $\bra{\psi^{(j)}} \psi^{(j+1)}\rangle = 0$ requires that 
\begin{equation}
    \sum_{n=1}^M \bar U_{1n}^{(j)} {U}_{1n}^{(j+1)} = 0.
\end{equation}
Furthermore, assuming without loss of generality that the system is initially prepared in the state $\ket{1}_1$, we set $U_{11}^{(j)} = 1/\sqrt[4]{5}$ to satisfy $\langle\psi^{(j)}|1\rangle_1 = 1/{\sqrt[4]{5}}$ for all $j$. Note that the remaining rows of each transfer matrix can then be chosen to ensure unitarity. This implies that each unitary $\mathcal{U}_j$ can be implemented using only $M-1$ beam splitters.

As an example, we consider the simplest case $M=3$, which is sufficient to demonstrate contextuality. In this case, we choose the unitary transformation $\mathcal{U}_j$ in Eq.~\eqref{Eq:Obs.A^{(j)}=UA0Ud} such that the first row of its transfer matrix is given by
\begin{equation}\label{Eq:Unitary}
    U_{1n}^{(j)} =
    \begin{cases}
        \cos\theta, & n=1, \\
        \sin\theta \cos\phi_j, & n=2, \\
        \sin\theta \sin\phi_j, & n=3.
    \end{cases}
\end{equation}
See Fig.~\ref{Fig:OpticalSetup}(b) and Appendix~\ref{appen:transfer-mat} for an explicit construction using two beam splitters. With this choice, the projectors in Eq.~\eqref{Eq:Obs.A^{(j)}=UA0Ud} are defined by the state vectors
\begin{equation}\label{Eq:Psi_j}
    \ket{\psi^{(j)}}
    = \cos\theta \ket{1}_1
    + \sin\theta \cos\phi_j \ket{1}_2
    + \sin\theta \sin\phi_j \ket{1}_3.
\end{equation}
By choosing $\cos\theta = 1/\sqrt[4]{5}$ and $\phi_j = 4\pi j/5$, we can see the conditions on the transfer matrices in Eq.~\eqref{Eq:Unitary} produce the set of optimal projectors described in Sec.~\ref{Sec:KCBS}, which achieve the maximal quantum violation.

As discussed in the previous section, the KCBS test can be expressed, via  Eq.~\eqref{Eq:Simp.KCBS}, in terms of the joint probabilities $p(v_j=+1, v_{j+1}=+1)$, which depend only on the $+1$ outcomes of sequential measurements. In our setup, this outcome corresponds to a no-click event, which does not destroy the single photon and allows access to the required joint probabilities. Using the sequential measurements defined above, we obtain $p(v_j=+1, v_{j+1}=+1)=1-\frac{2}{\sqrt{5}}$ and $p(v_j=+1)=1-\frac{1}{\sqrt{5}}$. Substituting these expressions into the KCBS inequality~\eqref{Eq:Simp.KCBS}, we find
\begin{equation}
    \mathcal{S}(\ket{1}_1)=\sqrt{5},
\end{equation}
which corresponds to the maximal quantum violation. This demonstrates that the above measurement configuration is optimal for an initial single-photon state in the first mode. Note that, since any single-photon state can be transformed into $\ket{1}_1$ via a suitable linear-optical unitary, combining such a transformation with the above measurement scheme allows one to construct optimal measurements for an arbitrary single-photon input state.

\section{Applications}\label{Sec:Verification}

In this section, we discuss applications of our scheme for witnessing the nonclassicality and photon-number statistics of a single-mode state injected into the first input port of our setup for the KCBS test. Specifically, we assume that the first mode is prepared in an arbitrary state $\rho$, while all remaining modes are in the vacuum state. The initial state of the $M$-mode system is therefore
\begin{equation}\label{Eq:GeneralInputState}
    \rho_1=\rho \otimes \ket{0}\!\bra{0}
    \otimes \cdots \otimes \ket{0}\!\bra{0}.
\end{equation}
In what follows, we develop a formalism to evaluate $\mathcal{S}(\rho_1)$ for a general input state~$\rho$ based on considering of no-click events from the on--off detectors in our setup. Note that, in general, the measurement projector $\Pi_-$ includes all photon-number components greater than or equal to one, whereas the projector corresponding to the no-click outcome remains $\Pi_+ = \ket{0}_1\!\bra{0}$, as in the previous section.

We begin by considering the case in which $\rho=\ket{\alpha}\!\bra{\alpha}$ is a coherent state. Using the measurement scheme defined in the previous section, specified by the unitary transformations in Eq.~\eqref{Eq:Unitary} with $\cos\theta = 1/\sqrt[4]{5}$ and $\phi_j = 4\pi j/5$, we obtain
\begin{equation}\label{Sq:S-CoherentState}
    \mathcal{S}(\ket{\alpha}_1)
    = 5\Big(e^{-|\alpha|^2/\sqrt{5}} - e^{-2|\alpha|^2/\sqrt{5}}\Big).
\end{equation}
A detailed derivation of this expression is provided in Appendix~\ref{appen:coherent-input}. As is clear from Eq.~\eqref{Eq:GeneralInputState}, throughout this section we use the notation $\ket{\phi}_1=\ket{\phi}\otimes \ket{0}\otimes\dots\otimes\ket{0}$, i.e., the subscript $1$ indicates that the state occupies the first mode, while all other modes are in the vacuum state.

One can simply check that $\mathcal{S}(\ket{\alpha}_1) \leq 1.25$, where the maximum value is attained at $|\alpha|^2 = \sqrt{5}\,\ln 2$. This bound has important implications for classical states. Any classical state $\rho_{\mathrm{cl}}$ can be viewed as a statistical mixture of coherent states using the Glauber-Sudarshan representation\cite{glauber1963photon,sudarshan1963equivalence}
\begin{equation}
    \rho_{\mathrm{cl}}
    = \int d^2\alpha \, P_{\mathrm{cl}}(\alpha)\,
    \proj{\alpha},
\end{equation}
where $P_{\mathrm{cl}}(\alpha)$ is a genuine probability density. If such a state is injected into the first input mode of our setup, we obtain
\begin{equation}
    \mathcal{S}(\rho_{\mathrm{cl},1})
    = \int d^2\alpha \, P_{\mathrm{cl}}(\alpha)\,
    \mathcal{S}(\ket{\alpha}_1)
    \leq 1.25.
\end{equation}
Here we have used the linearity of $\mathcal{S}$, namely $\mathcal{S}(X+Y)=\mathcal{S}(X)+\mathcal{S}(Y)$ for arbitrary operators $X$ and $Y$.

Therefore, observing a value $\mathcal{S}(\rho_1) > 1.25$ certifies the nonclassicality of the input state $\rho_1$. This improves the bound for witnessing the nonclassicality based on using the KCBS test in Ref.~\cite{zhang2019experimental}.

To proceed with the evaluation of $\mathcal{S}(\rho_1)$, we expand the single-mode state in the Fock basis, $\rho_1 = \sum_{nm}\rho_{nm}\ket{n}_{1}\!\bra{m}$, and the linearity of $\mc{S}(\rho_1)$ to write
\begin{equation}\label{Eq:Simp.KCBS4NState}
    \mathcal{S}(\rho_1) = \sum_{n,m = 0}^{\infty} \rho_{nm}  \mc{S}(\ket{n}_{1}\!\bra{m}).
\end{equation}
To evaluate $\mc{S}(\ket{n}_{1}\!\bra{m})$, we use the Glauber-Sudarshan representation for the operator $\ket{n}\!_{_1}\!\bra{m}$,
\begin{equation}\label{Eq:GlauberSudarshanRep}
    \ket{n}_{1}\!\bra{m} = \int d^2\alpha P_{nm}(\alpha) \ket{\alpha}_{1}\!\bra{\alpha}, 
\end{equation}
where the corresponding $P$-function is given by is given by \cite{rahimi2011quantum}
\begin{equation}\label{Eq:ProbN-State}
    P_{mn}(\alpha) = (-1)^{m+n} \frac{e^{|\alpha|^2}}{\sqrt{m!n!}} \partial^m_\alpha \partial^n_{\bar\alpha} \delta^2(\alpha),
\end{equation}
with $\partial^m_\alpha = \partial^m/\partial\alpha^m$, $\partial^n_{\bar\alpha} = \partial^n/\partial{\bar\alpha}^n$, $\delta^2(\alpha)=\delta(\alpha) \delta(\bar\alpha)$, and $\bar\alpha$ denotes the complex conjugate of $\alpha$. 

Substituting Eqs.~\eqref{Sq:S-CoherentState} and \eqref{Eq:GlauberSudarshanRep} into Eq.~\eqref{Eq:Simp.KCBS4NState}, and performing integration by parts, we obtain
\begin{equation}\label{Eq:Simp-KCBS4nmDensOper}
    \begin{split}
        \mc{S}(\ket{n}\hspace{-0.1em}_{1}\!\bra{m}) &= \frac{1}{\sqrt{m!n!}}\Big[\partial^n_{\bar\alpha}\partial^m_\alpha \big(e^{|\alpha|^2} \mathcal{S}(\ket{\alpha}_1) \big)\Big]\Big|_{|\alpha|=0} \\
        &=5\bigg(\!\Big(1-\frac{1}{\sqrt{5}}\Big)^{\!n}-\Big(1-\frac{2}{\sqrt{5}}\Big)^{\!n} \bigg)\delta_{nm}. 
    \end{split}
\end{equation}
Using this equation and Eq.~\eqref{Eq:Simp.KCBS4NState}, we can evaluate $\mathcal{S}(\rho_1)$ for an arbitrary single-mode state $\rho_1$. Note that Eq.~\eqref{Eq:Simp-KCBS4nmDensOper} shows that $\mathcal{S}(\rho_1)$ depends only on the diagonal elements $\rho_{nn}$ of the density operator.

To demonstrate the dependence of $\mathcal{S}(\rho_1)$ on the photon-number components of $\rho_1$, let us consider several examples. Using Eq.~\eqref{Eq:Simp-KCBS4nmDensOper}, we obtain $\mathcal{S}(\ket{0}_1)=0$, $\mathcal{S}(\ket{1}_1)=\sqrt{5}\approx 2.236$, $\mathcal{S}(\ket{2}_1)=2\sqrt{5}-3\approx 1.472$, $\mathcal{S}(\ket{3}_1)\approx 0.839$, and $\mathcal{S}(\ket{n}_1)<0.467$ for $n>3$. These values show that our setup can provide information about the photon-number components of an unknown input state. In particular, observing $\mathcal{S}(\rho_1)> 2\sqrt{5}-3 \approx 1.472$ implies that the single-photon population $\rho_{11}$ of the density operator is nonzero. Note that this method for verifying the presence of a single-photon component can be useful in situations where the standard $g^{(2)}$ test may fail. For example, if the state contains a single-photon contribution together with a small higher-order photon-number component, the $g^{(2)}$ criterion can fail to certify the presence of single-photon component \cite{reid1986violations}. Furthermore, observing $\mathcal{S}(\rho_1)\ge 0.839$ with our setup indicates the presence of single-photon and/or two-photon components in the state.

The above formalism also allows us to analyze the impact of experimental imperfections on the KCBS test discussed in the previous section. In particular, the presence of a vacuum component due to loss, or higher photon-number contributions arising from imperfect state preparation, generally reduces the observed value of $\mathcal{S}(\rho_1)$. As a first example, consider loss leading to a statistical mixture of vacuum and single-photon states,
$\rho_1 = \rho_{00}\ket{0}_1\!\bra{0}+ \rho_{11}\ket{1}_1\!\bra{1}$. Using the results above, we obtain $\mathcal{S}(\rho_1)= \rho_{11}\sqrt{5}$. Hence, a violation of the KCBS inequality, $\mathcal{S}(\rho_1)>2$, is still possible provided $\rho_{11} > \frac{2}{\sqrt{5}} \approx 0.894$. As a second example, consider a state containing single- and two-photon components, $\rho_1= \rho_{11}\ket{1}_1\!\bra{1} + \rho_{22}\ket{2}_1\!\bra{2}$.
In this case, $\mathcal{S}(\rho_1)= \rho_{11}\sqrt{5} + \rho_{22}(2\sqrt{5}-3)$. Imposing the violation condition $\mathcal{S}(\rho_1)>2$ yields $\rho_{11} > \frac{5-\sqrt{5}}{4}
\approx 0.691$. These examples illustrate quantitatively how vacuum and higher photon-number components diminish the observed contextuality violation.

\section{Experimental demonstration}\label{Sec:Experiment}

\begin{figure*}[t]
	\centering
	\includegraphics[width=\linewidth]{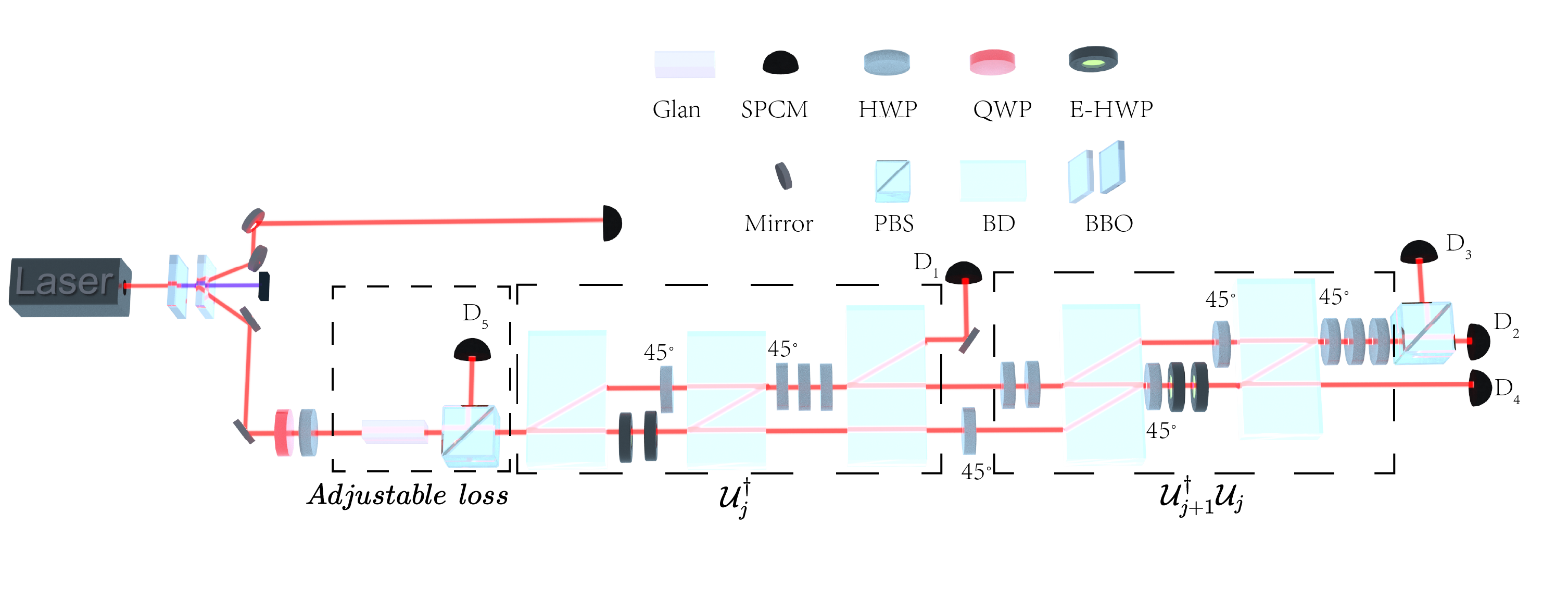}
	\caption{Experimental setup. A pulsed Ti:Sapphire laser is frequency doubled via parametric up-conversion in the first $\beta$-barium borate (BBO) crystal; the resulting second harmonic then pumps a second BBO crystal to generate photon pairs via spontaneous parametric down-conversion. One photon is routed through a Glan polarizer and a polarizing beam splitter (PBS) into the optical path, while the other (heralding photon) is directly coupled into a single-photon counting module (SPCM). The device within the dashed box is used to simulate photon loss in the optical path. The first three beam displacers (BDs) implement the unitary transformation $\mathcal U^{\dagger}_j$, and the subsequent two beam displacers realize the composite unitary operation $\mathcal U^{\dagger}_{j+1}\mathcal U_j$. Each of the four output ports of the optical circuit is coupled to an individual SPCM for detection and data processing.}
	\label{fig3}
\end{figure*}

We experimentally demonstrate our KCBS scheme using the setup illustrated in Fig.~\ref{fig3}. A pulsed Ti:sapphire laser (wavelength: 830 nm, repetition rate: 76 MHz) is frequency doubled in a BBO crystal. The resulting second harmonic (415 nm) then pumps a second BBO crystal to generate photon pairs via spontaneous parametric down-conversion. One photon from each pair passes through a quarter-wave plate, an HWP, a Glan polarizer and a polarizing beam splitter (PBS) to ensure horizontal polarization, while the other photon is directed to a single-photon counting module (SPCM) that serves as the heralding detector. The SPCM integrates an avalanche photodiode (APD) with a field-programmable gate array (FPGA) for synchronized detection.

To test the KCBS inequality, we implement three optical modes using a hybrid encoding of spatial paths and polarization. Specifically, the first mode corresponds to the up-path horizontal polarization ($\vert UH \rangle$), the second mode to the down-path horizontal polarization ($\vert DH \rangle$), and the third mode to the down-path vertical polarization ($\vert DV \rangle$). The sequential measurements are realized using five beam displacers (BDs) in combination with several half-wave plates.

As shown in Fig.~\ref{fig3}, two electrically controlled half-wave plates (E-HWPs) placed between the first and second BDs realize the first beam splitter operation in Fig.~\ref{Fig:OpticalSetup}(b). The combined transmission matrix of two cascaded half-wave plates (HWPs), $HWP(\theta_2) HWP(\theta_1)$, can be expressed as a rotation matrix that depends on the angle $2(\theta_1-\theta_2)$. Therefore, by setting $\theta_1-\theta_2 = \frac{\phi_j}{2}$, we realize the transmission matrix corresponding to the first beam splitter in Fig.~\ref{Fig:OpticalSetup}(b). Similarly, for the two half-wave plates following the $45^{\circ}$ HWP between the second and third BDs, we set $\theta_1-\theta_2 = \frac{\theta}{2}$ to implement the transmission matrix of the second beam splitter in Fig.~\ref{Fig:OpticalSetup}(b). The angle difference between the two HWPs in the upper path between the third and fourth BDs is set to $\theta_1-\theta_2 = -\frac{\theta}{2}$, corresponding to the transmission matrix of the third beam splitter in Fig.~\ref{Fig:OpticalSetup}(b). The two E-HWPs in the lower path between the fourth and fifth BDs are configured to realize the transformation
\begin{equation}
\begin{pmatrix}
\cos(\phi_{j+1}-\phi_j) & \sin(\phi_{j+1}-\phi_j) \\
-\sin(\phi_{j+1}-\phi_j) & \cos(\phi_{j+1}-\phi_j)
\end{pmatrix}.
\label{eq11}
\end{equation}
This transformation is obtained by setting the angular difference between the two E-HWPs to $(\phi_{j+1}-\phi_j)/2$. After the fifth BD, two additional HWPs placed in the upper path following a $45^{\circ}$ HWP are set with an angle difference of $\theta/2$. With this configuration, the combined action of the BDs and wave plates implements the theoretical scheme shown in Fig.~\ref{Fig:OpticalSetup}(b).

According to our scheme, a photon detection by $D_1$ corresponds to the event $v_j = -1$. A no-click event at $D_1$ followed by a photon detection at $D_2$ corresponds to $v_j = +1$ and $v_{j+1} = -1$. Similarly, a no-click event at $D_1$ together with a detection at either $D_3$ or $D_4$ corresponds to the joint event $v_j = +1$ and $v_{j+1} = +1$. As the probability of two detectors simultaneously registering photons (i.e., $p(v_j = -1,v_{j+1}= -1)$) is negligible, we can approximate $p(v_j = -1)\simeq p(v_j = -1,v_{j+1}= +1)$.

The Glan polarizer provides control over the polarization of transmitted light, and the subsequent PBS permits only horizontally polarized photons to be coupled into the main optical path. Through rotation of the Glan polarizer and monitoring of $D_5$, this configuration enables control over the vacuum field coupled into the path. 

Together, detector $D_5$, Glan polarizer, and PBS are used to simulate photon loss in the optical path, or in other words, to control the ratio of single-photon to vacuum components in the input state. If we do not monitor $D_5$ and instead consider only the events where a single photon is registered among $D_1$ to $D_4$, this corresponds  to the fair-sampling assumption. 

\section{Experimental Results}\label{Sec: Results}

The experimental results are shown in Tab.\ref{tab1}. We assume that the detected photons are statistically representative of all generated photons (fair-sampling assumption, when we do not monitor $D_5$). Our experimental results yield the KCBS value of $\mathcal S(\vert 1 \rangle_1) = 2.2363\pm0.0041$. The value of the KCBS quantity is calculated from the probability $P_{2,T}$. Each $P_{i,T}$ is obtained from the coincidence count rates $C_{j,T}$ between detector j and the trigger detector. For example, $P_{1,T} = \frac{C_{1,T}}{\sum C_{j,T}}$. For each measurement cycle, detection events are accumulated over a $1$~s interval, during which approximately $4100$ coincidence counts are typically recorded. Each experimental conjfiguration is repeated $100$ times, and the results are averaged. The standard deviation of the mean is used to determine the statistical uncertainties. The experimental results demonstrate a clear violation of the KCBS inequality~\ref{Eq:Simp.KCBS}. Furthermore, our sequential measurement scheme resolves the issue of non-identical compatible measurements encountered in earlier experiments~\cite{zhang2019experimental,lapkiewicz2011experimental}.
\begin{table*}[ht]
	\centering
	\begin{tabular}{ccccccccc}

		\hline\hline\noalign{\smallskip}	
		    & $C_{1,T}$& $C_{2,T}$ & $C_{3,T}$& $C_{4,T}$ & $P_{1,T}$& $P_{2,T}$ & $P_{3,T}$& $P_{4,T}$   \\
		\noalign{\smallskip}\hline\noalign{\smallskip}
		$\{A^{(1)}, A^{(2)}\}$ & 4073(248) & 4070(250) & 179(17)&769(57)&0.4480(0.0022)&0.4477(0.0020)&0.0197(0.0015)&0.0846(0.0027) \\
		$\{A^{(2)}, A^{(3)}\}$ & 4294(236) & 4283(237) & 185(16)&814(52)&0.4484(0.0025)&0.4472(0.0020)&0.0193(0.0013)&0.0850(0.0024) \\
		$\{A^{(3)}, A^{(4)}\}$ & 4647(261) &4641(266) &203(22)&887(57)&0.4478(0.0021)&0.4472(0.0017)&0.0196(0.0015)&0.0855(0.0019)\\
		$\{A^{(4)}, A^{(5)}\}$ & 4604(275) & 4595(272) & 204(19)&876(56)&0.4478(0.0022)&0.4470(0.0016)&0.0198(0.0013)&0.0853(0.0023) \\
		$\{A^{(5)}, A^{(1)}\}$ & 4146(250) & 4142(255) & 187(17)&787(52)&0.4477(0.0017)&0.4472(0.0019)&0.0202(0.0015)&0.0850(0.0023)\\
		\noalign{\smallskip}\hline
	\end{tabular}
    	\caption{The registered measurement results. Here $C_{j,T}$ denotes the coincidence count rate in 1s between the detectors $1,(2,3,4)$ and the trigger detector in idler mode. $P_{i,T}$ represents the probabilities that are used to evaluate the joint probabilities. The values in the brackets represent standard uncertainties given by 100 repetitions of the experiment.} 
       \label{tab1} 
\end{table*}

\begin{figure}[t]
	\centering
	\includegraphics[width=\linewidth]{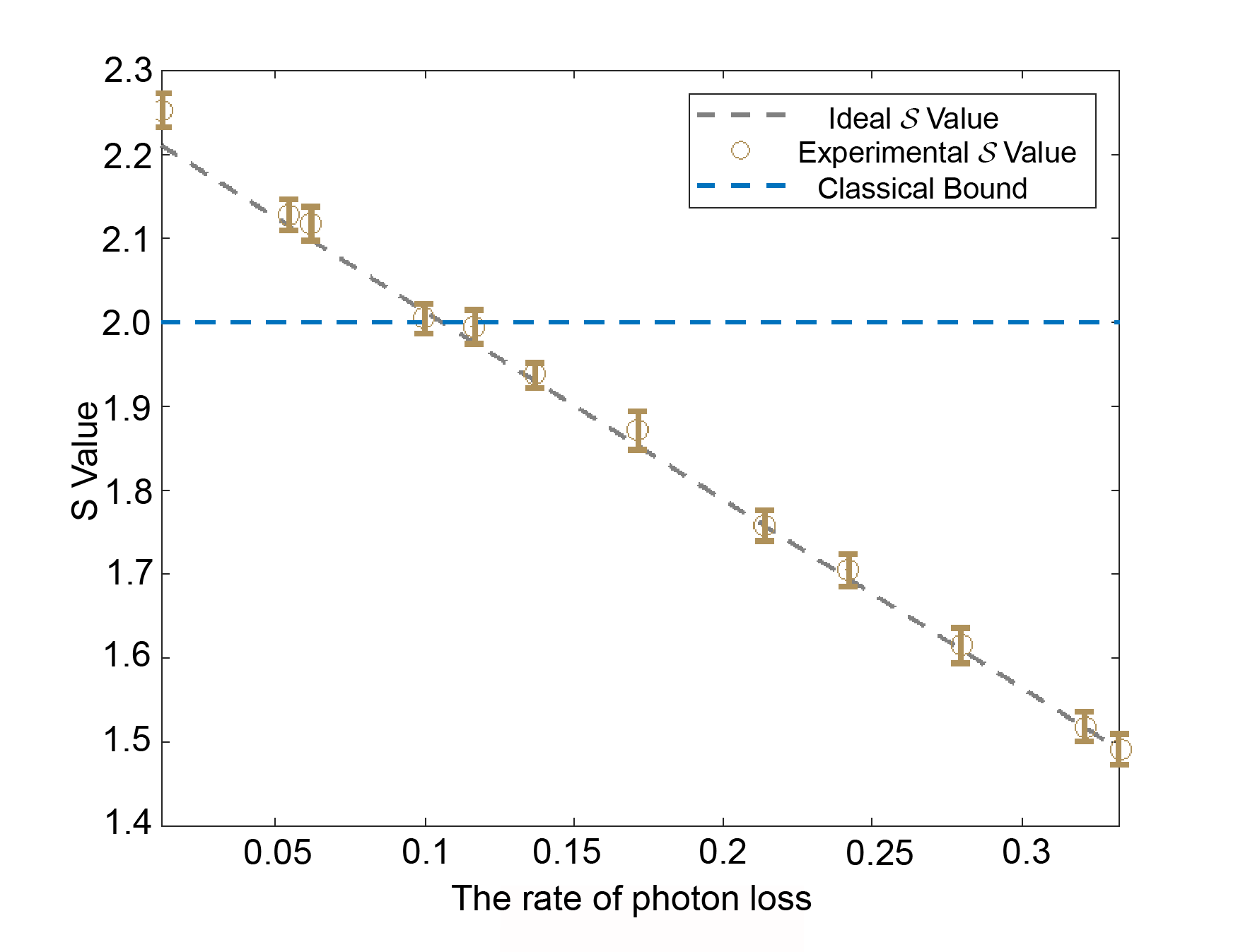}
	\caption{The extent of violation of the KCBS inequality is quantified by how much the KCBS value $\mathcal{S}$ exceeds the classical bound of $2$. The dependence of this quantity on the photon-loss rate is shown, where values above and below $2$ correspond to violation and adherence to the inequality, respectively. Here, the ``Ideal $\mathcal{S}$ Value'' denotes the theoretical prediction $\mathcal{S}(\rho_1)$ for the input state $\rho_1=\rho_{00}\ket{\bm0}\!\bra{\bm0}+\rho_{11}\ket{1}_1\!\bra{1}$, where $\rho_{00}$ represents the photon-loss rate. The ``Experimental $\mathcal{S}$ Value'' corresponds to the measured data obtained in our experiment. The ``Classical Bound'' indicates the threshold value in the inequality that separates the classical and quantum regimes.}
	\label{fig4}
\end{figure}

  We also investigate the effect of photon loss on the KCBS inequality violation by monitoring detector $D_5$. In the presence of loss, the input state to our test can be modeled as $\rho_1=\rho_{00}\ket{\bm0}\!\bra{\bm0}+\rho_{11}\ket{1}_1\!\bra{1}$, where $\rho_{00}$ denotes the vacuum component and can be viewed as the photon-loss rate. Figure~\ref{fig4} shows the ideal and experimental values of the KCBS quantity $\mathcal{S}(\rho_1)$ as a function of the photon-loss rate. Violations of the inequality (values $>2$) are observed for low loss rates, while for larger losses the value drops below $2$, indicating that the inequality is no longer violated.
  
 These results demonstrate that the observed value of the KCBS quantity $\mathcal{S}$ can be used to verify the presence of the single-photon component in the experimental state. In particular, the threshold value $\mathcal{S}=2$ corresponds to a single-photon fraction $\rho_{11}=\frac{2}{\sqrt{5}}\approx 0.894$. As discussed in Sec.~\ref{Sec:Verification}, observing a value $\mathcal{S}>1.472$ always certifies the presence of a nonzero single-photon component. Compared with the conventional quantum optics criterion $g^{(2)}$, a significant advantage of our approach is that while $g^{(2)}$ becomes ill-defined when the input state contains a vacuum component, our method still yields meaningful physical results.
  
\section{conclusion and outlook}\label{Sec:Conclu}

In this work, we design and implement a linear-optical demonstration of Kochen--Specker contextuality based on the violation of the KCBS inequality, which must be satisfied by any noncontextual value assignment. Our setup realizes genuinely sequential measurements using linear-optical networks and on--off photodetection. A key feature of our scheme is the use of vacuum (no-click) detection events, which allows the sequential measurements to satisfy the essential requirements of repeatability and non-disturbance. This approach simplifies experimental implementation while fulfilling the core Kochen--Specker requirements for operational or black-box verifications of contextuality \cite{cabello2016simple}.

In our construction, measurements are implemented such that every co-measured observable within a given context remains physically identical across different contexts. This resolves a central limitation of previous experimental demonstrations \cite{zhang2019experimental,lapkiewicz2011experimental}, where compatible measurements were not realized by identical physical operations.

Beyond demonstrating contextuality, our scheme also enables practical applications. In particular, it can be used to verify nonclassicality and probe the photon-number composition of an arbitrary single-mode state. Our analysis further shows that the scheme remains robust in the presence of photon loss. These results indicate that, under a trusted-device assumption, our setup can serve as a practical tool for verifying single-photon sources. The experimental results exhibit a clear violation of the KCBS inequality, providing strong evidence against noncontextual hidden-variable descriptions of the observed correlations.

More broadly, the proposed approach offers a versatile framework for characterizing photonic quantum systems and the linear-optical networks that process them. The ability to infer photon-number statistics and certify properties of input states and measurement devices suggests applications in the self-testing and certification of photonic components, including single-photon sources, detectors, and interferometric networks used in modern quantum technologies. At a more fundamental level, the contextuality revealed by our scheme reflects the interplay between the wave and particle aspects of single photons. The particle-like behavior is manifested in the mutual exclusivity of detector clicks, while the wave-like behavior arises from the coherent superposition of a single photon across different modes, which leads to the violation of the KCBS inequality. This connection points to potential links between contextuality tests and other interferometric probes of quantum coherence and metrological advantage \cite{catani,Wagner2024coherence}. Future work may therefore explore contextuality-based protocols for certifying coherence resources, enhancing interferometric sensing, and developing device-independent or semi-device-independent verification methods for photonic quantum systems.


\section*{ACKNOWLEDGMENTS}

S.R.-K. acknowledges useful discussions with M. Rezai. S.R-K. acknowledges support from the Knut and Alice Wallenberg Foundation through the Wallenberg Center for Quantum Technology (WACQT). This work was supported by the National Natural Science Foundation of China (Nos. 12347104, U24A2017, and 12461160276), the National Key Research and Development Program of China (No. 2023YFC2205802), and the Natural Science Foundation of Jiangsu Province (Nos.BK20243060 and BK20233001). B.O. acknowledges the \textit{Ernst Mach Grant - worldwide}, administered by OeAD and financed by the Federal Ministry Women, Science and Research Republic of Austria (BMFWF). B.O. also acknowledges financial support from the Austrian Science Fund (FWF) under Projects [\href{https://doi.org/10.55776/P35810}{10.55776/P35810}] and [\href{https://doi.org/10.55776/P36633}{10.55776/P36633}].

\appendix
\section{Connection Between the KCBS Inequality and Its Simplified Form} \label{appen.KCBS}

We derive the connection between the KCBS inequality in Eq.~\eqref{eq.KCBS} and its simplified form in Eq.~\eqref{Eq:Simp.KCBS}. Since the observables $A^{(j)}$ and $A^{(j+1)}$ commute by construction, their associated projectors commute as well. Exploiting this property, the joint probabilities defined in Eq.~\eqref{eq:joint-prob} can be rewritten as
\begin{equation}
     p(v_j, v_{j+1})=\tr \big[\rho \,\Pi^{(j)}_{v_j} \Pi^{(j+1)}_{v_{j+1}} \big],
\end{equation}
where we have used the cyclicity of the trace together with the idempotency of the projectors,
$\Pi^{(j)}_{v_j}\Pi^{(j)}_{v_j}=\Pi^{(j)}_{v_j}$.
Using this expression and the relation $\Pi^{(j)}_{+} =(\mathbb{1}+ A^{(j)})/2$, the simplified KCBS quantity in Eq.~\eqref{Eq:Simp.KCBS} can be written as
\begin{align}
        \mathcal{S}(\rho)
        &= \sum_{j=1}^{5} p(v_j=+1) - \sum_{j=1}^{5} p(v_j=+1,v_{j+1}=+1)\nonumber\\
        &= \sum_{j=1}^{5}\Big( \tr \big[\rho \Pi^{(j)}_+ \big]
        - \tr \big[\rho \Pi^{(j)}_+ \Pi^{(j+1)}_+ \big]\Big)\nonumber\\
        &= \sum_{j=1}^{5}\frac{1}{4}\Big( 1
        + \tr \big[\rho \big(A^{(j)}\!-A^{(j+1)}\big) \big]
        {-} \tr \big[\rho A^{(j)} A^{(j+1)} \big]\Big)\nonumber\\
        &= \sum_{j=1}^{5}\frac{1}{4}\Big( 1
        - \big\langle A^{(j)} A^{(j+1)} \big\rangle \Big)\nonumber\\
        &= \frac{5-\kappa(\rho)}{4}.
\end{align}
In the third line, the term
$\sum_{j=1}^{5}\tr \big[\rho \big(A^{(j)}-A^{(j+1)}\big) \big]$
vanishes due to the cyclic convention $A^{(6)} \equiv A^{(1)}$, which ensures a telescopic cancellation over the sum.


\section{The transfer matrix for $M=3$}\label{appen:transfer-mat}

The transfer matrix corresponding to a beam splitter acting on modes $j$ and $k$ is given by
\begin{equation}\label{eq:BS}
    B_{\omega,jk}=
    \begin{pmatrix}
        \cos \omega & \sin \omega \\
        -\sin \omega & \cos \omega
    \end{pmatrix},
\end{equation}
where $\cos \omega$ denotes the amplitude transmissivity from the $j$th input mode to the $j$th output mode. If a single photon is injected into the $j$th input mode while the $k$th mode is in the vacuum state, the probability of detecting the photon in the $j$th output mode is $\cos^{2}\omega$.

For the case $M=3$, as shown in Fig.~\ref{Fig:OpticalSetup}(b), the linear-optical network before and after the on--off photodetector can be constructed using only two beam splitters. Using Eq.~\eqref{eq:BS}, the corresponding transfer matrix is
\begin{align}
    U^{(j)} &= (B_{\theta,12}\oplus I_1)(I_1\oplus B_{\phi_j,12}) \nonumber \\
    &=
    \begin{pmatrix}
        \cos\theta & \sin\theta \cos\phi_j & \sin\theta \sin\phi_j \\
        -\sin\theta & \cos\theta \cos\phi_j & \cos\theta \sin\phi_j \\
        0 & -\sin\phi_j & \cos\phi_j
    \end{pmatrix},
\end{align}
where $I_1$ denotes the $1\times1$ identity matrix.
 

\section{Derivation of $\mathcal{S}(\ket{\alpha}_1)$}\label{appen:coherent-input}

In this appendix, we derive the expression for $\mathcal{S}(\ket{\alpha}_1)$ given in Eq.~\eqref{Sq:S-CoherentState}. We begin by showing that linear-optical networks transform coherent states into coherent states.

The $M$-mode displacement operator is defined as
\begin{align}
D(\bm{\alpha})
= \exp\!\left(\bm{\alpha}\bm{a}^\dagger - \bm{a}\bm{\alpha}^\dagger\right),
\end{align}
where $\bm{\alpha}=(\alpha_1,\alpha_2,\ldots,\alpha_M)$, 
$\bm{a}=(a_1,\ldots,a_M)$ is the row vector of the annihilation operators, and 
$\bm{a}^\dagger=(a_1^\dagger,\ldots,a_M^\dagger)^{T}$ is the column vector of the creation operators. Using this notation, Eq.~\eqref{Eq:LON}, which defines a lossless linear-optical network, can be written compactly as
\begin{equation}
    \mathcal{U}\,\bm{a}^\dagger\,\mathcal{U}^\dagger
    = U\,\bm{a}^\dagger,
\end{equation}
where $U$ is the $M\times M$ unitary transfer matrix describing the network.

Applying a linear-optical network to an $M$-mode coherent state $\ket{\bm{\alpha}}=\ket{\alpha_1,\alpha_2,\ldots,\alpha_M}$ yields
\begin{align}
\mathcal{U}\ket{\bm{\alpha}}
&= \mathcal{U} D(\bm{\alpha}) \mathcal{U}^\dagger \ket{\bm{0}} \nonumber \\
&= \exp\!\left(\bm{\alpha} U \bm{a}^\dagger - \bm{a} U^\dagger \bm{\alpha}^\dagger\right)\ket{\bm{0}} \nonumber \\
&= \ket{\bm{\alpha} U},
\end{align}
where $\ket{\bm{0}}$ denotes the $M$-mode vacuum state. Similarly, one finds
\begin{equation}
    \mathcal{U}^\dagger \ket{\bm{\alpha}}
    = \ket{\bm{\alpha} U^\dagger}.
\end{equation}
For further details on the properties of linear-optical networks, see Ref.~\cite{rahimi2020}.

Using these relations, we can compute the probabilities of detecting a no-click outcome from the on--off detector when a coherent state $\ket{\alpha}$ is injected into the first input mode of our setup (Fig.~\ref{Fig:OpticalSetup}) instead of the single-photon state. After detecting a no-click outcome from the first measurement, the post-measurement state is proportional to an unnormalized coherent state
\begin{align}
    \ket{\tilde{\bm\mu}^{(j)}}
    &= \mathcal{U}_j \Big(\proj{0}\otimes I \otimes\dots\otimes I\Big)\mathcal{U}_j^\dagger
       \ket{\alpha,0,\dots,0} \nonumber\\
    &= \mathcal{U}_j \Big(\proj{0}\otimes I \otimes\dots\otimes I\Big)
       \big|\bm\beta^{(j)}\big\rangle \nonumber\\
    &= \exp\!\Big(-\tfrac12 \big|\beta_1^{(j)}\big|^2\Big)
       \mathcal{U}_j \big|\bm\gamma^{(j)}\big\rangle \nonumber\\
    &= \exp\!\Big(-\tfrac12 |\alpha|^2 \big|U_{11}^{(j)}\big|^2\Big)
       \big|\bm\gamma^{(j)} U^{(j)}\big\rangle,
\end{align}
where 
\[
\bm\beta^{(j)}
= \big(\alpha \bar{U}_{11}^{(j)}, \alpha \bar{U}_{21}^{(j)},\dots,\alpha \bar{U}_{M1}^{(j)}\big),
\]
and
\[
\bm\gamma^{(j)}
= \big(0, \alpha \bar{U}_{21}^{(j)},\dots,\alpha \bar{U}_{M1}^{(j)}\big).
\]
Using the orthogonality relation between rows of unitary matrices, $\sum_{n=1}^M \bar{U}_{n1}^{(j)} U_{nk}^{(j)}=\delta_{1k}$, we obtain
\begin{align}
    \bm\gamma^{(j)} U^{(j)}
    &= \Big(\alpha \sum_{n=2}^M \big|U_{n1}^{(j)}\big|^2,
        \dots,
        \alpha \sum_{n=2}^M \bar{U}_{n1}^{(j)} U_{nM}^{(j)}\Big) \\
    &= \alpha \Big(
        1-\big|U_{11}^{(j)}\big|^2,
        -\bar{U}_{11}^{(j)} U_{12}^{(j)},
        \dots,
        -\bar{U}_{11}^{(j)} U_{1M}^{(j)}
        \Big),\nonumber
\end{align}
and
\begin{equation}
    \big|\bm\gamma^{(j)} U^{(j)}\big|^2
    = \bm\gamma^{(j)} {\bm\gamma^{(j)}}^\dagger
    = 1 - \big|U_{11}^{(j)}\big|^2 .
\end{equation}

Using these expressions, the probability of detecting a no-click outcome for measurement $(j)$ is
\begin{align}\label{eq:coh-no-click}
    p(v_j=+1)
    &= \big\langle \alpha,0,\dots,0 \big| \tilde{\bm\mu}^{(j)} \big\rangle \nonumber\\
    &= \exp\!\Big(-|\alpha|^2 |U_{11}^{(j)}|^2\Big).
\end{align}

We can also compute the joint probability of detecting no-clicks in the sequential measurements $(j)$ and $(j+1)$ in our setup,
\begin{align}
    p(v_j=+1,v_{j+1}&=+1)=\big\langle \tilde{\bm\mu}^{(j+1)}\big| \tilde{\bm\mu}^{(j)}\big\rangle\\
    =\exp\!\bigg(&  |\alpha|^2\Big(1- \big|{U}_{11}^{(j)}\big|^2\Big) \Big(1- \big|{U}_{11}^{(j+1)}\big|^2\Big)\nonumber\\
    &-|\alpha|^2+{U}_{11}^{(j)} \bar{U}_{11}^{(j+1)} \sum_{n=2}^M{U}_{1n}^{(j)} \bar{U}_{1n}^{(j+1)}\bigg),\nonumber
\end{align}
which depends on the first row of the unitary transfer matrices. If we choose the first rows of the transfer matrices $(j)$ and $(j+1)$ to be orthogonal,
\[
\sum_{n=1}^M U_{1n}^{(j)} \bar{U}_{1n}^{(j+1)} = 0,
\]
as required for the KCBS test, and set $U_{11}^{(j)} = 1/\sqrt[4]{5}$ as discussed in Sec.~\ref{Sec:Methods}, we obtain
\begin{equation}
    p(v_j=+1,v_{j+1}=+1)
    =\exp\!\Big(\!-\frac{2}{\sqrt{5}}|\alpha|^2\Big).
\end{equation}

Substituting this expression together with Eq.~\eqref{eq:coh-no-click} into Eq.~\eqref{Eq:Simp.KCBS}, we obtain the KCBS value for a coherent state $\ket{\alpha}$ injected into the first input mode of our setup,
\begin{equation}
 \mathcal{S}(\ket{\alpha}_1)
    = 5\Big(e^{-|\alpha|^2/\sqrt{5}} - e^{-2|\alpha|^2/\sqrt{5}}\Big).
\end{equation}
Note that in this scenario the joint probability $p(v_j=+1,v_{j+1}=+1)$ is independent of the measurement order.


\begin{thebibliography}{54}%
	\makeatletter
	\providecommand \@ifxundefined [1]{%
		\@ifx{#1\undefined}
	}%
	\providecommand \@ifnum [1]{%
		\ifnum #1\expandafter \@firstoftwo
		\else \expandafter \@secondoftwo
		\fi
	}%
	\providecommand \@ifx [1]{%
		\ifx #1\expandafter \@firstoftwo
		\else \expandafter \@secondoftwo
		\fi
	}%
	\providecommand \natexlab [1]{#1}%
	\providecommand \enquote  [1]{``#1''}%
	\providecommand \bibnamefont  [1]{#1}%
	\providecommand \bibfnamefont [1]{#1}%
	\providecommand \citenamefont [1]{#1}%
	\providecommand \href@noop [0]{\@secondoftwo}%
	\providecommand \href [0]{\begingroup \@sanitize@url \@href}%
	\providecommand \@href[1]{\@@startlink{#1}\@@href}%
	\providecommand \@@href[1]{\endgroup#1\@@endlink}%
	\providecommand \@sanitize@url [0]{\catcode `\\12\catcode `\$12\catcode
		`\&12\catcode `\#12\catcode `\^12\catcode `\_12\catcode `\%12\relax}%
	\providecommand \@@startlink[1]{}%
	\providecommand \@@endlink[0]{}%
	\providecommand \url  [0]{\begingroup\@sanitize@url \@url }%
	\providecommand \@url [1]{\endgroup\@href {#1}{\urlprefix }}%
	\providecommand \urlprefix  [0]{URL }%
	\providecommand \Eprint [0]{\href }%
	\providecommand \doibase [0]{https://doi.org/}%
	\providecommand \selectlanguage [0]{\@gobble}%
	\providecommand \bibinfo  [0]{\@secondoftwo}%
	\providecommand \bibfield  [0]{\@secondoftwo}%
	\providecommand \translation [1]{[#1]}%
	\providecommand \BibitemOpen [0]{}%
	\providecommand \bibitemStop [0]{}%
	\providecommand \bibitemNoStop [0]{.\EOS\space}%
	\providecommand \EOS [0]{\spacefactor3000\relax}%
	\providecommand \BibitemShut  [1]{\csname bibitem#1\endcsname}%
	\let\auto@bib@innerbib\@empty
	\bibitem [{\citenamefont {Kochen}\ and\ \citenamefont
		{Specker}(1967)}]{kochen1967the}%
	\BibitemOpen
	\bibfield  {author} {\bibinfo {author} {\bibfnamefont {S.}~\bibnamefont
			{Kochen}}\ and\ \bibinfo {author} {\bibfnamefont {E.~P.}\ \bibnamefont
			{Specker}},\ }\bibfield  {title} {\bibinfo {title} {The problem of hidden
			variables in quantum mechanics},\ }\href
	{http://www.jstor.org/stable/24902153} {\bibfield  {journal} {\bibinfo
			{journal} {Journal of Mathematics and Mechanics}\ }\textbf {\bibinfo {volume}
			{17}},\ \bibinfo {pages} {59} (\bibinfo {year} {1967})}\BibitemShut {NoStop}%
	\bibitem [{\citenamefont {BELL}(1966)}]{bell1966problem}%
	\BibitemOpen
	\bibfield  {author} {\bibinfo {author} {\bibfnamefont {J.~S.}\ \bibnamefont
			{BELL}},\ }\bibfield  {title} {\bibinfo {title} {On the problem of hidden
			variables in quantum mechanics},\ }\href
	{https://doi.org/10.1103/RevModPhys.38.447} {\bibfield  {journal} {\bibinfo
			{journal} {Rev. Mod. Phys.}\ }\textbf {\bibinfo {volume} {38}},\ \bibinfo
		{pages} {447} (\bibinfo {year} {1966})}\BibitemShut {NoStop}%
	\bibitem [{\citenamefont {Budroni}\ \emph {et~al.}(2022)\citenamefont
		{Budroni}, \citenamefont {Cabello}, \citenamefont {G\"uhne}, \citenamefont
		{Kleinmann},\ and\ \citenamefont {Larsson}}]{budroni2022kochen}%
	\BibitemOpen
	\bibfield  {author} {\bibinfo {author} {\bibfnamefont {C.}~\bibnamefont
			{Budroni}}, \bibinfo {author} {\bibfnamefont {A.}~\bibnamefont {Cabello}},
		\bibinfo {author} {\bibfnamefont {O.}~\bibnamefont {G\"uhne}}, \bibinfo
		{author} {\bibfnamefont {M.}~\bibnamefont {Kleinmann}},\ and\ \bibinfo
		{author} {\bibfnamefont {J.-A.}\ \bibnamefont {Larsson}},\ }\bibfield
	{title} {\bibinfo {title} {Kochen-specker contextuality},\ }\href
	{https://doi.org/10.1103/RevModPhys.94.045007} {\bibfield  {journal}
		{\bibinfo  {journal} {Rev. Mod. Phys.}\ }\textbf {\bibinfo {volume} {94}},\
		\bibinfo {pages} {045007} (\bibinfo {year} {2022})}\BibitemShut {NoStop}%
	\bibitem [{\citenamefont {Specker}(1960)}]{Specker1960-SPEDLN}%
	\BibitemOpen
	\bibfield  {author} {\bibinfo {author} {\bibfnamefont {E.}~\bibnamefont
			{Specker}},\ }\bibfield  {title} {\bibinfo {title} {Die logik nicht
			gleichzeitig entsc heidbarer aussagen},\ }\href
	{https://doi.org/10.1111/j.1746-8361.1960.tb00422.x} {\bibfield  {journal}
		{\bibinfo  {journal} {Dialectica}\ }\textbf {\bibinfo {volume} {14}},\
		\bibinfo {pages} {239} (\bibinfo {year} {1960})}\BibitemShut {NoStop}%
	\bibitem [{\citenamefont {Brunner}\ \emph {et~al.}(2014)\citenamefont
		{Brunner}, \citenamefont {Cavalcanti}, \citenamefont {Pironio}, \citenamefont
		{Scarani},\ and\ \citenamefont {Wehner}}]{Nonlocality}%
	\BibitemOpen
	\bibfield  {author} {\bibinfo {author} {\bibfnamefont {N.}~\bibnamefont
			{Brunner}}, \bibinfo {author} {\bibfnamefont {D.}~\bibnamefont {Cavalcanti}},
		\bibinfo {author} {\bibfnamefont {S.}~\bibnamefont {Pironio}}, \bibinfo
		{author} {\bibfnamefont {V.}~\bibnamefont {Scarani}},\ and\ \bibinfo {author}
		{\bibfnamefont {S.}~\bibnamefont {Wehner}},\ }\bibfield  {title} {\bibinfo
		{title} {Bell nonlocality},\ }\href
	{https://doi.org/10.1103/RevModPhys.86.419} {\bibfield  {journal} {\bibinfo
			{journal} {Rev. Mod. Phys.}\ }\textbf {\bibinfo {volume} {86}},\ \bibinfo
		{pages} {419} (\bibinfo {year} {2014})}\BibitemShut {NoStop}%
	\bibitem [{\citenamefont {Nagata}(2005)}]{nagata2005kochen}%
	\BibitemOpen
	\bibfield  {author} {\bibinfo {author} {\bibfnamefont {K.}~\bibnamefont
			{Nagata}},\ }\bibfield  {title} {\bibinfo {title} {Kochen-specker theorem as
			a precondition for secure quantum key distribution},\ }\href
	{https://doi.org/10.1103/PhysRevA.72.012325} {\bibfield  {journal} {\bibinfo
			{journal} {Phys. Rev. A}\ }\textbf {\bibinfo {volume} {72}},\ \bibinfo
		{pages} {012325} (\bibinfo {year} {2005})}\BibitemShut {NoStop}%
	\bibitem [{\citenamefont {Saha}\ and\ \citenamefont
		{Chaturvedi}(2019)}]{saha2019preparation}%
	\BibitemOpen
	\bibfield  {author} {\bibinfo {author} {\bibfnamefont {D.}~\bibnamefont
			{Saha}}\ and\ \bibinfo {author} {\bibfnamefont {A.}~\bibnamefont
			{Chaturvedi}},\ }\bibfield  {title} {\bibinfo {title} {Preparation
			contextuality as an essential feature underlying quantum communication
			advantage},\ }\href {https://doi.org/10.1103/PhysRevA.100.022108} {\bibfield
		{journal} {\bibinfo  {journal} {Phys. Rev. A}\ }\textbf {\bibinfo {volume}
			{100}},\ \bibinfo {pages} {022108} (\bibinfo {year} {2019})}\BibitemShut
	{NoStop}%
	\bibitem [{\citenamefont {Gupta}\ \emph {et~al.}(2023)\citenamefont {Gupta},
		\citenamefont {Saha}, \citenamefont {Xu}, \citenamefont {Cabello},\ and\
		\citenamefont {Majumdar}}]{gupta2023quantum}%
	\BibitemOpen
	\bibfield  {author} {\bibinfo {author} {\bibfnamefont {S.}~\bibnamefont
			{Gupta}}, \bibinfo {author} {\bibfnamefont {D.}~\bibnamefont {Saha}},
		\bibinfo {author} {\bibfnamefont {Z.-P.}\ \bibnamefont {Xu}}, \bibinfo
		{author} {\bibfnamefont {A.}~\bibnamefont {Cabello}},\ and\ \bibinfo {author}
		{\bibfnamefont {A.~S.}\ \bibnamefont {Majumdar}},\ }\bibfield  {title}
	{\bibinfo {title} {Quantum contextuality provides communication complexity
			advantage},\ }\href {https://doi.org/10.1103/PhysRevLett.130.080802}
	{\bibfield  {journal} {\bibinfo  {journal} {Phys. Rev. Lett.}\ }\textbf
		{\bibinfo {volume} {130}},\ \bibinfo {pages} {080802} (\bibinfo {year}
		{2023})}\BibitemShut {NoStop}%
	\bibitem [{\citenamefont {Schmid}\ and\ \citenamefont
		{Spekkens}(2018)}]{PRXschmid}%
	\BibitemOpen
	\bibfield  {author} {\bibinfo {author} {\bibfnamefont {D.}~\bibnamefont
			{Schmid}}\ and\ \bibinfo {author} {\bibfnamefont {R.~W.}\ \bibnamefont
			{Spekkens}},\ }\bibfield  {title} {\bibinfo {title} {Contextual advantage for
			state discrimination},\ }\href {https://doi.org/10.1103/PhysRevX.8.011015}
	{\bibfield  {journal} {\bibinfo  {journal} {Phys. Rev. X}\ }\textbf {\bibinfo
			{volume} {8}},\ \bibinfo {pages} {011015} (\bibinfo {year}
		{2018})}\BibitemShut {NoStop}%
	\bibitem [{\citenamefont {Howard}\ \emph {et~al.}(2014)\citenamefont {Howard},
		\citenamefont {Wallman}, \citenamefont {Veitch},\ and\ \citenamefont
		{Emerson}}]{howard2014contextuality}%
	\BibitemOpen
	\bibfield  {author} {\bibinfo {author} {\bibfnamefont {M.}~\bibnamefont
			{Howard}}, \bibinfo {author} {\bibfnamefont {J.}~\bibnamefont {Wallman}},
		\bibinfo {author} {\bibfnamefont {V.}~\bibnamefont {Veitch}},\ and\ \bibinfo
		{author} {\bibfnamefont {J.}~\bibnamefont {Emerson}},\ }\bibfield  {title}
	{\bibinfo {title} {Contextuality supplies the `magic' for quantum
			computation},\ }\href {https://doi.org/10.1038/nature13460} {\bibfield
		{journal} {\bibinfo  {journal} {Nature}\ }\textbf {\bibinfo {volume} {510}},\
		\bibinfo {pages} {351} (\bibinfo {year} {2014})}\BibitemShut {NoStop}%
	\bibitem [{\citenamefont {Delfosse}\ \emph {et~al.}(2015)\citenamefont
		{Delfosse}, \citenamefont {Allard~Guerin}, \citenamefont {Bian},\ and\
		\citenamefont {Raussendorf}}]{delfosse2015wigner}%
	\BibitemOpen
	\bibfield  {author} {\bibinfo {author} {\bibfnamefont {N.}~\bibnamefont
			{Delfosse}}, \bibinfo {author} {\bibfnamefont {P.}~\bibnamefont
			{Allard~Guerin}}, \bibinfo {author} {\bibfnamefont {J.}~\bibnamefont
			{Bian}},\ and\ \bibinfo {author} {\bibfnamefont {R.}~\bibnamefont
			{Raussendorf}},\ }\bibfield  {title} {\bibinfo {title} {Wigner function
			negativity and contextuality in quantum computation on rebits},\ }\href
	{https://doi.org/10.1103/PhysRevX.5.021003} {\bibfield  {journal} {\bibinfo
			{journal} {Phys. Rev. X}\ }\textbf {\bibinfo {volume} {5}},\ \bibinfo {pages}
		{021003} (\bibinfo {year} {2015})}\BibitemShut {NoStop}%
	\bibitem [{\citenamefont {Dawkins}\ and\ \citenamefont
		{Howard}(2015)}]{dawkins2015qutrit}%
	\BibitemOpen
	\bibfield  {author} {\bibinfo {author} {\bibfnamefont {H.}~\bibnamefont
			{Dawkins}}\ and\ \bibinfo {author} {\bibfnamefont {M.}~\bibnamefont
			{Howard}},\ }\bibfield  {title} {\bibinfo {title} {Qutrit magic state
			distillation tight in some directions},\ }\href
	{https://doi.org/10.1103/PhysRevLett.115.030501} {\bibfield  {journal}
		{\bibinfo  {journal} {Phys. Rev. Lett.}\ }\textbf {\bibinfo {volume} {115}},\
		\bibinfo {pages} {030501} (\bibinfo {year} {2015})}\BibitemShut {NoStop}%
	\bibitem [{\citenamefont {Raussendorf}(2013)}]{raussendorf2013contextuality}%
	\BibitemOpen
	\bibfield  {author} {\bibinfo {author} {\bibfnamefont {R.}~\bibnamefont
			{Raussendorf}},\ }\bibfield  {title} {\bibinfo {title} {Contextuality in
			measurement-based quantum computation},\ }\href
	{https://doi.org/10.1103/PhysRevA.88.022322} {\bibfield  {journal} {\bibinfo
			{journal} {Phys. Rev. A}\ }\textbf {\bibinfo {volume} {88}},\ \bibinfo
		{pages} {022322} (\bibinfo {year} {2013})}\BibitemShut {NoStop}%
	\bibitem [{\citenamefont {Bermejo-Vega}\ \emph {et~al.}(2017)\citenamefont
		{Bermejo-Vega}, \citenamefont {Delfosse}, \citenamefont {Browne},
		\citenamefont {Okay},\ and\ \citenamefont
		{Raussendorf}}]{bermejo2017contextuality}%
	\BibitemOpen
	\bibfield  {author} {\bibinfo {author} {\bibfnamefont {J.}~\bibnamefont
			{Bermejo-Vega}}, \bibinfo {author} {\bibfnamefont {N.}~\bibnamefont
			{Delfosse}}, \bibinfo {author} {\bibfnamefont {D.~E.}\ \bibnamefont
			{Browne}}, \bibinfo {author} {\bibfnamefont {C.}~\bibnamefont {Okay}},\ and\
		\bibinfo {author} {\bibfnamefont {R.}~\bibnamefont {Raussendorf}},\
	}\bibfield  {title} {\bibinfo {title} {Contextuality as a resource for models
			of quantum computation with qubits},\ }\href
	{https://doi.org/10.1103/PhysRevLett.119.120505} {\bibfield  {journal}
		{\bibinfo  {journal} {Phys. Rev. Lett.}\ }\textbf {\bibinfo {volume} {119}},\
		\bibinfo {pages} {120505} (\bibinfo {year} {2017})}\BibitemShut {NoStop}%
	\bibitem [{\citenamefont {Raussendorf}\ \emph {et~al.}(2017)\citenamefont
		{Raussendorf}, \citenamefont {Browne}, \citenamefont {Delfosse},
		\citenamefont {Okay},\ and\ \citenamefont
		{Bermejo-Vega}}]{raussendorf2017contextuality}%
	\BibitemOpen
	\bibfield  {author} {\bibinfo {author} {\bibfnamefont {R.}~\bibnamefont
			{Raussendorf}}, \bibinfo {author} {\bibfnamefont {D.~E.}\ \bibnamefont
			{Browne}}, \bibinfo {author} {\bibfnamefont {N.}~\bibnamefont {Delfosse}},
		\bibinfo {author} {\bibfnamefont {C.}~\bibnamefont {Okay}},\ and\ \bibinfo
		{author} {\bibfnamefont {J.}~\bibnamefont {Bermejo-Vega}},\ }\bibfield
	{title} {\bibinfo {title} {Contextuality and wigner-function negativity in
			qubit quantum computation},\ }\href
	{https://doi.org/10.1103/PhysRevA.95.052334} {\bibfield  {journal} {\bibinfo
			{journal} {Phys. Rev. A}\ }\textbf {\bibinfo {volume} {95}},\ \bibinfo
		{pages} {052334} (\bibinfo {year} {2017})}\BibitemShut {NoStop}%
	\bibitem [{\citenamefont {G\"uhne}\ \emph {et~al.}(2014)\citenamefont
		{G\"uhne}, \citenamefont {Budroni}, \citenamefont {Cabello}, \citenamefont
		{Kleinmann},\ and\ \citenamefont {Larsson}}]{dimensionWitness}%
	\BibitemOpen
	\bibfield  {author} {\bibinfo {author} {\bibfnamefont {O.}~\bibnamefont
			{G\"uhne}}, \bibinfo {author} {\bibfnamefont {C.}~\bibnamefont {Budroni}},
		\bibinfo {author} {\bibfnamefont {A.}~\bibnamefont {Cabello}}, \bibinfo
		{author} {\bibfnamefont {M.}~\bibnamefont {Kleinmann}},\ and\ \bibinfo
		{author} {\bibfnamefont {J.-A.}\ \bibnamefont {Larsson}},\ }\bibfield
	{title} {\bibinfo {title} {Bounding the quantum dimension with
			contextuality},\ }\href {https://doi.org/10.1103/PhysRevA.89.062107}
	{\bibfield  {journal} {\bibinfo  {journal} {Phys. Rev. A}\ }\textbf {\bibinfo
			{volume} {89}},\ \bibinfo {pages} {062107} (\bibinfo {year}
		{2014})}\BibitemShut {NoStop}%
	\bibitem [{\citenamefont {Bharti}\ \emph {et~al.}(2019)\citenamefont {Bharti},
		\citenamefont {Ray}, \citenamefont {Varvitsiotis}, \citenamefont {Warsi},
		\citenamefont {Cabello},\ and\ \citenamefont {Kwek}}]{SelfTesting}%
	\BibitemOpen
	\bibfield  {author} {\bibinfo {author} {\bibfnamefont {K.}~\bibnamefont
			{Bharti}}, \bibinfo {author} {\bibfnamefont {M.}~\bibnamefont {Ray}},
		\bibinfo {author} {\bibfnamefont {A.}~\bibnamefont {Varvitsiotis}}, \bibinfo
		{author} {\bibfnamefont {N.~A.}\ \bibnamefont {Warsi}}, \bibinfo {author}
		{\bibfnamefont {A.}~\bibnamefont {Cabello}},\ and\ \bibinfo {author}
		{\bibfnamefont {L.-C.}\ \bibnamefont {Kwek}},\ }\bibfield  {title} {\bibinfo
		{title} {Robust self-testing of quantum systems via noncontextuality
			inequalities},\ }\href {https://doi.org/10.1103/PhysRevLett.122.250403}
	{\bibfield  {journal} {\bibinfo  {journal} {Phys. Rev. Lett.}\ }\textbf
		{\bibinfo {volume} {122}},\ \bibinfo {pages} {250403} (\bibinfo {year}
		{2019})}\BibitemShut {NoStop}%
	\bibitem [{\citenamefont {Klyachko}\ \emph {et~al.}(2008)\citenamefont
		{Klyachko}, \citenamefont {Can}, \citenamefont
		{Binicio\ifmmode~\breve{g}\else \u{g}\fi{}lu},\ and\ \citenamefont
		{Shumovsky}}]{klyachko2008simple}%
	\BibitemOpen
	\bibfield  {author} {\bibinfo {author} {\bibfnamefont {A.~A.}\ \bibnamefont
			{Klyachko}}, \bibinfo {author} {\bibfnamefont {M.~A.}\ \bibnamefont {Can}},
		\bibinfo {author} {\bibfnamefont {S.}~\bibnamefont
			{Binicio\ifmmode~\breve{g}\else \u{g}\fi{}lu}},\ and\ \bibinfo {author}
		{\bibfnamefont {A.~S.}\ \bibnamefont {Shumovsky}},\ }\bibfield  {title}
	{\bibinfo {title} {Simple test for hidden variables in spin-1 systems},\
	}\href {https://doi.org/10.1103/PhysRevLett.101.020403} {\bibfield  {journal}
		{\bibinfo  {journal} {Phys. Rev. Lett.}\ }\textbf {\bibinfo {volume} {101}},\
		\bibinfo {pages} {020403} (\bibinfo {year} {2008})}\BibitemShut {NoStop}%
	\bibitem [{\citenamefont {Yu}\ and\ \citenamefont {Oh}(2012)}]{yu2012state}%
	\BibitemOpen
	\bibfield  {author} {\bibinfo {author} {\bibfnamefont {S.}~\bibnamefont
			{Yu}}\ and\ \bibinfo {author} {\bibfnamefont {C.~H.}\ \bibnamefont {Oh}},\
	}\bibfield  {title} {\bibinfo {title} {State-independent proof of
			kochen-specker theorem with 13 rays},\ }\href
	{https://doi.org/10.1103/PhysRevLett.108.030402} {\bibfield  {journal}
		{\bibinfo  {journal} {Phys. Rev. Lett.}\ }\textbf {\bibinfo {volume} {108}},\
		\bibinfo {pages} {030402} (\bibinfo {year} {2012})}\BibitemShut {NoStop}%
	\bibitem [{\citenamefont {Clauser}\ \emph {et~al.}(1969)\citenamefont
		{Clauser}, \citenamefont {Horne}, \citenamefont {Shimony},\ and\
		\citenamefont {Holt}}]{clauser1969proposed}%
	\BibitemOpen
	\bibfield  {author} {\bibinfo {author} {\bibfnamefont {J.~F.}\ \bibnamefont
			{Clauser}}, \bibinfo {author} {\bibfnamefont {M.~A.}\ \bibnamefont {Horne}},
		\bibinfo {author} {\bibfnamefont {A.}~\bibnamefont {Shimony}},\ and\ \bibinfo
		{author} {\bibfnamefont {R.~A.}\ \bibnamefont {Holt}},\ }\bibfield  {title}
	{\bibinfo {title} {Proposed experiment to test local hidden-variable
			theories},\ }\href {https://doi.org/10.1103/PhysRevLett.23.880} {\bibfield
		{journal} {\bibinfo  {journal} {Phys. Rev. Lett.}\ }\textbf {\bibinfo
			{volume} {23}},\ \bibinfo {pages} {880} (\bibinfo {year} {1969})}\BibitemShut
	{NoStop}%
	\bibitem [{\citenamefont {Mermin}(1990)}]{mermin1990simple}%
	\BibitemOpen
	\bibfield  {author} {\bibinfo {author} {\bibfnamefont {N.~D.}\ \bibnamefont
			{Mermin}},\ }\bibfield  {title} {\bibinfo {title} {Simple unified form for
			the major no-hidden-variables theorems},\ }\href
	{https://doi.org/10.1103/PhysRevLett.65.3373} {\bibfield  {journal} {\bibinfo
			{journal} {Phys. Rev. Lett.}\ }\textbf {\bibinfo {volume} {65}},\ \bibinfo
		{pages} {3373} (\bibinfo {year} {1990})}\BibitemShut {NoStop}%
	\bibitem [{\citenamefont {Mermin}(1993)}]{mermin1993hidden}%
	\BibitemOpen
	\bibfield  {author} {\bibinfo {author} {\bibfnamefont {N.~D.}\ \bibnamefont
			{Mermin}},\ }\bibfield  {title} {\bibinfo {title} {Hidden variables and the
			two theorems of john bell},\ }\href
	{https://doi.org/10.1103/RevModPhys.65.803} {\bibfield  {journal} {\bibinfo
			{journal} {Rev. Mod. Phys.}\ }\textbf {\bibinfo {volume} {65}},\ \bibinfo
		{pages} {803} (\bibinfo {year} {1993})}\BibitemShut {NoStop}%
	\bibitem [{\citenamefont {Peres}(1990)}]{peres1990incompatible}%
	\BibitemOpen
	\bibfield  {author} {\bibinfo {author} {\bibfnamefont {A.}~\bibnamefont
			{Peres}},\ }\bibfield  {title} {\bibinfo {title} {Incompatible results of
			quantum measurements},\ }\href
	{https://www.sciencedirect.com/science/article/pii/037596019090172K}
	{\bibfield  {journal} {\bibinfo  {journal} {Physics Letters A}\ }\textbf
		{\bibinfo {volume} {151}},\ \bibinfo {pages} {107} (\bibinfo {year}
		{1990})}\BibitemShut {NoStop}%
	\bibitem [{\citenamefont {Peres}(1991)}]{peres1991two}%
	\BibitemOpen
	\bibfield  {author} {\bibinfo {author} {\bibfnamefont {A.}~\bibnamefont
			{Peres}},\ }\bibfield  {title} {\bibinfo {title} {Two simple proofs of the
			kochen-specker theorem},\ }\href {https://doi.org/10.1088/0305-4470/24/4/003}
	{\bibfield  {journal} {\bibinfo  {journal} {Journal of Physics A:
				Mathematical and General}\ }\textbf {\bibinfo {volume} {24}},\ \bibinfo
		{pages} {L175} (\bibinfo {year} {1991})}\BibitemShut {NoStop}%
	\bibitem [{\citenamefont {Lapkiewicz}\ \emph {et~al.}(2011)\citenamefont
		{Lapkiewicz}, \citenamefont {Li}, \citenamefont {Schaeff}, \citenamefont
		{Langford}, \citenamefont {Ramelow}, \citenamefont {Wie{\'{s}}niak},\ and\
		\citenamefont {Zeilinger}}]{lapkiewicz2011experimental}%
	\BibitemOpen
	\bibfield  {author} {\bibinfo {author} {\bibfnamefont {R.}~\bibnamefont
			{Lapkiewicz}}, \bibinfo {author} {\bibfnamefont {P.}~\bibnamefont {Li}},
		\bibinfo {author} {\bibfnamefont {C.}~\bibnamefont {Schaeff}}, \bibinfo
		{author} {\bibfnamefont {N.~K.}\ \bibnamefont {Langford}}, \bibinfo {author}
		{\bibfnamefont {S.}~\bibnamefont {Ramelow}}, \bibinfo {author} {\bibfnamefont
			{M.}~\bibnamefont {Wie{\'{s}}niak}},\ and\ \bibinfo {author} {\bibfnamefont
			{A.}~\bibnamefont {Zeilinger}},\ }\bibfield  {title} {\bibinfo {title}
		{Experimental non-classicality of an indivisible quantum system},\ }\href
	{https://doi.org/10.1038/nature10119} {\bibfield  {journal} {\bibinfo
			{journal} {Nature}\ }\textbf {\bibinfo {volume} {474}},\ \bibinfo {pages}
		{490} (\bibinfo {year} {2011})}\BibitemShut {NoStop}%
	\bibitem [{\citenamefont {Kirchmair}\ \emph {et~al.}(2009)\citenamefont
		{Kirchmair}, \citenamefont {Z{\"a}hringer}, \citenamefont {Gerritsma},
		\citenamefont {Kleinmann}, \citenamefont {G{\"u}hne}, \citenamefont
		{Cabello}, \citenamefont {Blatt},\ and\ \citenamefont
		{Roos}}]{kirchmair2009state}%
	\BibitemOpen
	\bibfield  {author} {\bibinfo {author} {\bibfnamefont {G.}~\bibnamefont
			{Kirchmair}}, \bibinfo {author} {\bibfnamefont {F.}~\bibnamefont
			{Z{\"a}hringer}}, \bibinfo {author} {\bibfnamefont {R.}~\bibnamefont
			{Gerritsma}}, \bibinfo {author} {\bibfnamefont {M.}~\bibnamefont
			{Kleinmann}}, \bibinfo {author} {\bibfnamefont {O.}~\bibnamefont
			{G{\"u}hne}}, \bibinfo {author} {\bibfnamefont {A.}~\bibnamefont {Cabello}},
		\bibinfo {author} {\bibfnamefont {R.}~\bibnamefont {Blatt}},\ and\ \bibinfo
		{author} {\bibfnamefont {C.~F.}\ \bibnamefont {Roos}},\ }\bibfield  {title}
	{\bibinfo {title} {State-independent experimental test of quantum
			contextuality},\ }\href {https://doi.org/10.1038/nature08172} {\bibfield
		{journal} {\bibinfo  {journal} {Nature}\ }\textbf {\bibinfo {volume} {460}},\
		\bibinfo {pages} {494} (\bibinfo {year} {2009})}\BibitemShut {NoStop}%
	\bibitem [{\citenamefont {Bartosik}\ \emph {et~al.}(2009)\citenamefont
		{Bartosik}, \citenamefont {Klepp}, \citenamefont {Schmitzer}, \citenamefont
		{Sponar}, \citenamefont {Cabello}, \citenamefont {Rauch},\ and\ \citenamefont
		{Hasegawa}}]{bartosik2009experimental}%
	\BibitemOpen
	\bibfield  {author} {\bibinfo {author} {\bibfnamefont {H.}~\bibnamefont
			{Bartosik}}, \bibinfo {author} {\bibfnamefont {J.}~\bibnamefont {Klepp}},
		\bibinfo {author} {\bibfnamefont {C.}~\bibnamefont {Schmitzer}}, \bibinfo
		{author} {\bibfnamefont {S.}~\bibnamefont {Sponar}}, \bibinfo {author}
		{\bibfnamefont {A.}~\bibnamefont {Cabello}}, \bibinfo {author} {\bibfnamefont
			{H.}~\bibnamefont {Rauch}},\ and\ \bibinfo {author} {\bibfnamefont
			{Y.}~\bibnamefont {Hasegawa}},\ }\bibfield  {title} {\bibinfo {title}
		{Experimental test of quantum contextuality in neutron interferometry},\
	}\href {https://doi.org/10.1103/PhysRevLett.103.040403} {\bibfield  {journal}
		{\bibinfo  {journal} {Phys. Rev. Lett.}\ }\textbf {\bibinfo {volume} {103}},\
		\bibinfo {pages} {040403} (\bibinfo {year} {2009})}\BibitemShut {NoStop}%
	\bibitem [{\citenamefont {Jerger}\ \emph {et~al.}(2016)\citenamefont {Jerger},
		\citenamefont {Reshitnyk}, \citenamefont {Oppliger}, \citenamefont
		{Poto{\v{c}}nik}, \citenamefont {Mondal}, \citenamefont {Wallraff},
		\citenamefont {Goodenough}, \citenamefont {Wehner}, \citenamefont
		{Juliusson}, \citenamefont {Langford},\ and\ \citenamefont
		{Fedorov}}]{jerger2016contextuality}%
	\BibitemOpen
	\bibfield  {author} {\bibinfo {author} {\bibfnamefont {M.}~\bibnamefont
			{Jerger}}, \bibinfo {author} {\bibfnamefont {Y.}~\bibnamefont {Reshitnyk}},
		\bibinfo {author} {\bibfnamefont {M.}~\bibnamefont {Oppliger}}, \bibinfo
		{author} {\bibfnamefont {A.}~\bibnamefont {Poto{\v{c}}nik}}, \bibinfo
		{author} {\bibfnamefont {M.}~\bibnamefont {Mondal}}, \bibinfo {author}
		{\bibfnamefont {A.}~\bibnamefont {Wallraff}}, \bibinfo {author}
		{\bibfnamefont {K.}~\bibnamefont {Goodenough}}, \bibinfo {author}
		{\bibfnamefont {S.}~\bibnamefont {Wehner}}, \bibinfo {author} {\bibfnamefont
			{K.}~\bibnamefont {Juliusson}}, \bibinfo {author} {\bibfnamefont {N.~K.}\
			\bibnamefont {Langford}},\ and\ \bibinfo {author} {\bibfnamefont
			{A.}~\bibnamefont {Fedorov}},\ }\bibfield  {title} {\bibinfo {title}
		{Contextuality without nonlocality in a superconducting quantum system},\
	}\href {https://doi.org/10.1038/ncomms12930} {\bibfield  {journal} {\bibinfo
			{journal} {Nature Communications}\ }\textbf {\bibinfo {volume} {7}},\
		\bibinfo {pages} {12930} (\bibinfo {year} {2016})}\BibitemShut {NoStop}%
	\bibitem [{\citenamefont {Huang}\ \emph {et~al.}(2003)\citenamefont {Huang},
		\citenamefont {Li}, \citenamefont {Zhang}, \citenamefont {Pan},\ and\
		\citenamefont {Guo}}]{huang2003experimental}%
	\BibitemOpen
	\bibfield  {author} {\bibinfo {author} {\bibfnamefont {Y.-F.}\ \bibnamefont
			{Huang}}, \bibinfo {author} {\bibfnamefont {C.-F.}\ \bibnamefont {Li}},
		\bibinfo {author} {\bibfnamefont {Y.-S.}\ \bibnamefont {Zhang}}, \bibinfo
		{author} {\bibfnamefont {J.-W.}\ \bibnamefont {Pan}},\ and\ \bibinfo {author}
		{\bibfnamefont {G.-C.}\ \bibnamefont {Guo}},\ }\bibfield  {title} {\bibinfo
		{title} {Experimental test of the kochen-specker theorem with single
			photons},\ }\href {https://doi.org/10.1103/PhysRevLett.90.250401} {\bibfield
		{journal} {\bibinfo  {journal} {Phys. Rev. Lett.}\ }\textbf {\bibinfo
			{volume} {90}},\ \bibinfo {pages} {250401} (\bibinfo {year}
		{2003})}\BibitemShut {NoStop}%
	\bibitem [{\citenamefont {Amselem}\ \emph {et~al.}(2009)\citenamefont
		{Amselem}, \citenamefont {R\aa{}dmark}, \citenamefont {Bourennane},\ and\
		\citenamefont {Cabello}}]{SICphoton}%
	\BibitemOpen
	\bibfield  {author} {\bibinfo {author} {\bibfnamefont {E.}~\bibnamefont
			{Amselem}}, \bibinfo {author} {\bibfnamefont {M.}~\bibnamefont
			{R\aa{}dmark}}, \bibinfo {author} {\bibfnamefont {M.}~\bibnamefont
			{Bourennane}},\ and\ \bibinfo {author} {\bibfnamefont {A.}~\bibnamefont
			{Cabello}},\ }\bibfield  {title} {\bibinfo {title} {State-independent quantum
			contextuality with single photons},\ }\href
	{https://doi.org/10.1103/PhysRevLett.103.160405} {\bibfield  {journal}
		{\bibinfo  {journal} {Phys. Rev. Lett.}\ }\textbf {\bibinfo {volume} {103}},\
		\bibinfo {pages} {160405} (\bibinfo {year} {2009})}\BibitemShut {NoStop}%
	\bibitem [{\citenamefont {Ahrens}\ \emph {et~al.}(2013)\citenamefont {Ahrens},
		\citenamefont {Amselem}, \citenamefont {Cabello},\ and\ \citenamefont
		{Bourennane}}]{ahrens2013two}%
	\BibitemOpen
	\bibfield  {author} {\bibinfo {author} {\bibfnamefont {J.}~\bibnamefont
			{Ahrens}}, \bibinfo {author} {\bibfnamefont {E.}~\bibnamefont {Amselem}},
		\bibinfo {author} {\bibfnamefont {A.}~\bibnamefont {Cabello}},\ and\ \bibinfo
		{author} {\bibfnamefont {M.}~\bibnamefont {Bourennane}},\ }\bibfield  {title}
	{\bibinfo {title} {Two fundamental experimental tests of nonclassicality with
			qutrits},\ }\href {https://doi.org/10.1038/srep02170} {\bibfield  {journal}
		{\bibinfo  {journal} {Scientific Reports}\ }\textbf {\bibinfo {volume} {3}},\
		\bibinfo {pages} {2170} (\bibinfo {year} {2013})}\BibitemShut {NoStop}%
	\bibitem [{\citenamefont {D'Ambrosio}\ \emph {et~al.}(2013)\citenamefont
		{D'Ambrosio}, \citenamefont {Herbauts}, \citenamefont {Amselem},
		\citenamefont {Nagali}, \citenamefont {Bourennane}, \citenamefont
		{Sciarrino},\ and\ \citenamefont {Cabello}}]{d2013experimental}%
	\BibitemOpen
	\bibfield  {author} {\bibinfo {author} {\bibfnamefont {V.}~\bibnamefont
			{D'Ambrosio}}, \bibinfo {author} {\bibfnamefont {I.}~\bibnamefont
			{Herbauts}}, \bibinfo {author} {\bibfnamefont {E.}~\bibnamefont {Amselem}},
		\bibinfo {author} {\bibfnamefont {E.}~\bibnamefont {Nagali}}, \bibinfo
		{author} {\bibfnamefont {M.}~\bibnamefont {Bourennane}}, \bibinfo {author}
		{\bibfnamefont {F.}~\bibnamefont {Sciarrino}},\ and\ \bibinfo {author}
		{\bibfnamefont {A.}~\bibnamefont {Cabello}},\ }\bibfield  {title} {\bibinfo
		{title} {Experimental implementation of a kochen-specker set of quantum
			tests},\ }\href {https://doi.org/10.1103/PhysRevX.3.011012} {\bibfield
		{journal} {\bibinfo  {journal} {Phys. Rev. X}\ }\textbf {\bibinfo {volume}
			{3}},\ \bibinfo {pages} {011012} (\bibinfo {year} {2013})}\BibitemShut
	{NoStop}%
	\bibitem [{\citenamefont {Amselem}\ \emph {et~al.}(2012)\citenamefont
		{Amselem}, \citenamefont {Danielsen}, \citenamefont {L\'opez-Tarrida},
		\citenamefont {Portillo}, \citenamefont {Bourennane},\ and\ \citenamefont
		{Cabello}}]{amselem2012experimental}%
	\BibitemOpen
	\bibfield  {author} {\bibinfo {author} {\bibfnamefont {E.}~\bibnamefont
			{Amselem}}, \bibinfo {author} {\bibfnamefont {L.~E.}\ \bibnamefont
			{Danielsen}}, \bibinfo {author} {\bibfnamefont {A.~J.}\ \bibnamefont
			{L\'opez-Tarrida}}, \bibinfo {author} {\bibfnamefont {J.~R.}\ \bibnamefont
			{Portillo}}, \bibinfo {author} {\bibfnamefont {M.}~\bibnamefont
			{Bourennane}},\ and\ \bibinfo {author} {\bibfnamefont {A.}~\bibnamefont
			{Cabello}},\ }\bibfield  {title} {\bibinfo {title} {Experimental fully
			contextual correlations},\ }\href
	{https://doi.org/10.1103/PhysRevLett.108.200405} {\bibfield  {journal}
		{\bibinfo  {journal} {Phys. Rev. Lett.}\ }\textbf {\bibinfo {volume} {108}},\
		\bibinfo {pages} {200405} (\bibinfo {year} {2012})}\BibitemShut {NoStop}%
	\bibitem [{\citenamefont {Arias}\ \emph {et~al.}(2015)\citenamefont {Arias},
		\citenamefont {Ca\~nas}, \citenamefont {G\'omez}, \citenamefont {Barra},
		\citenamefont {Xavier}, \citenamefont {Lima}, \citenamefont {D'Ambrosio},
		\citenamefont {Baccari}, \citenamefont {Sciarrino},\ and\ \citenamefont
		{Cabello}}]{arias2015testing}%
	\BibitemOpen
	\bibfield  {author} {\bibinfo {author} {\bibfnamefont {M.}~\bibnamefont
			{Arias}}, \bibinfo {author} {\bibfnamefont {G.}~\bibnamefont {Ca\~nas}},
		\bibinfo {author} {\bibfnamefont {E.~S.}\ \bibnamefont {G\'omez}}, \bibinfo
		{author} {\bibfnamefont {J.~F.}\ \bibnamefont {Barra}}, \bibinfo {author}
		{\bibfnamefont {G.~B.}\ \bibnamefont {Xavier}}, \bibinfo {author}
		{\bibfnamefont {G.}~\bibnamefont {Lima}}, \bibinfo {author} {\bibfnamefont
			{V.}~\bibnamefont {D'Ambrosio}}, \bibinfo {author} {\bibfnamefont
			{F.}~\bibnamefont {Baccari}}, \bibinfo {author} {\bibfnamefont
			{F.}~\bibnamefont {Sciarrino}},\ and\ \bibinfo {author} {\bibfnamefont
			{A.}~\bibnamefont {Cabello}},\ }\bibfield  {title} {\bibinfo {title} {Testing
			noncontextuality inequalities that are building blocks of quantum
			correlations},\ }\href {https://doi.org/10.1103/PhysRevA.92.032126}
	{\bibfield  {journal} {\bibinfo  {journal} {Phys. Rev. A}\ }\textbf {\bibinfo
			{volume} {92}},\ \bibinfo {pages} {032126} (\bibinfo {year}
		{2015})}\BibitemShut {NoStop}%
	\bibitem [{\citenamefont {Zu}\ \emph {et~al.}(2012)\citenamefont {Zu},
		\citenamefont {Wang}, \citenamefont {Deng}, \citenamefont {Chang},
		\citenamefont {Liu}, \citenamefont {Hou}, \citenamefont {Yang},\ and\
		\citenamefont {Duan}}]{zu2012state}%
	\BibitemOpen
	\bibfield  {author} {\bibinfo {author} {\bibfnamefont {C.}~\bibnamefont
			{Zu}}, \bibinfo {author} {\bibfnamefont {Y.-X.}\ \bibnamefont {Wang}},
		\bibinfo {author} {\bibfnamefont {D.-L.}\ \bibnamefont {Deng}}, \bibinfo
		{author} {\bibfnamefont {X.-Y.}\ \bibnamefont {Chang}}, \bibinfo {author}
		{\bibfnamefont {K.}~\bibnamefont {Liu}}, \bibinfo {author} {\bibfnamefont
			{P.-Y.}\ \bibnamefont {Hou}}, \bibinfo {author} {\bibfnamefont {H.-X.}\
			\bibnamefont {Yang}},\ and\ \bibinfo {author} {\bibfnamefont {L.-M.}\
			\bibnamefont {Duan}},\ }\bibfield  {title} {\bibinfo {title}
		{State-independent experimental test of quantum contextuality in an
			indivisible system},\ }\href {https://doi.org/10.1103/PhysRevLett.109.150401}
	{\bibfield  {journal} {\bibinfo  {journal} {Phys. Rev. Lett.}\ }\textbf
		{\bibinfo {volume} {109}},\ \bibinfo {pages} {150401} (\bibinfo {year}
		{2012})}\BibitemShut {NoStop}%
	\bibitem [{\citenamefont {Borges}\ \emph {et~al.}(2014)\citenamefont {Borges},
		\citenamefont {Carvalho}, \citenamefont {de~Assis}, \citenamefont {Ferraz},
		\citenamefont {Ara\'ujo}, \citenamefont {Cabello}, \citenamefont {Cunha},\
		and\ \citenamefont {P\'adua}}]{borges2014quantum}%
	\BibitemOpen
	\bibfield  {author} {\bibinfo {author} {\bibfnamefont {G.}~\bibnamefont
			{Borges}}, \bibinfo {author} {\bibfnamefont {M.}~\bibnamefont {Carvalho}},
		\bibinfo {author} {\bibfnamefont {P.-L.}\ \bibnamefont {de~Assis}}, \bibinfo
		{author} {\bibfnamefont {J.}~\bibnamefont {Ferraz}}, \bibinfo {author}
		{\bibfnamefont {M.}~\bibnamefont {Ara\'ujo}}, \bibinfo {author}
		{\bibfnamefont {A.}~\bibnamefont {Cabello}}, \bibinfo {author} {\bibfnamefont
			{M.~T.}\ \bibnamefont {Cunha}},\ and\ \bibinfo {author} {\bibfnamefont
			{S.~a.}\ \bibnamefont {P\'adua}},\ }\bibfield  {title} {\bibinfo {title}
		{Quantum contextuality in a young-type interference experiment},\ }\href
	{https://doi.org/10.1103/PhysRevA.89.052106} {\bibfield  {journal} {\bibinfo
			{journal} {Phys. Rev. A}\ }\textbf {\bibinfo {volume} {89}},\ \bibinfo
		{pages} {052106} (\bibinfo {year} {2014})}\BibitemShut {NoStop}%
	\bibitem [{\citenamefont {Asadian}\ and\ \citenamefont
		{Cabello}(2022)}]{asadian2022bosonic}%
	\BibitemOpen
	\bibfield  {author} {\bibinfo {author} {\bibfnamefont {A.}~\bibnamefont
			{Asadian}}\ and\ \bibinfo {author} {\bibfnamefont {A.}~\bibnamefont
			{Cabello}},\ }\bibfield  {title} {\bibinfo {title} {Bosonic
			indistinguishability-dependent contextuality},\ }\href
	{https://doi.org/10.1103/PhysRevA.105.012404} {\bibfield  {journal} {\bibinfo
			{journal} {Phys. Rev. A}\ }\textbf {\bibinfo {volume} {105}},\ \bibinfo
		{pages} {012404} (\bibinfo {year} {2022})}\BibitemShut {NoStop}%
	\bibitem [{\citenamefont {Frustaglia}\ \emph {et~al.}(2016)\citenamefont
		{Frustaglia}, \citenamefont {Baltan\'as}, \citenamefont
		{Vel\'azquez-Ahumada}, \citenamefont {Fern\'andez-Prieto}, \citenamefont
		{Lujambio}, \citenamefont {Losada}, \citenamefont {Freire},\ and\
		\citenamefont {Cabello}}]{Clcorr}%
	\BibitemOpen
	\bibfield  {author} {\bibinfo {author} {\bibfnamefont {D.}~\bibnamefont
			{Frustaglia}}, \bibinfo {author} {\bibfnamefont {J.~P.}\ \bibnamefont
			{Baltan\'as}}, \bibinfo {author} {\bibfnamefont {M.~C.}\ \bibnamefont
			{Vel\'azquez-Ahumada}}, \bibinfo {author} {\bibfnamefont {A.}~\bibnamefont
			{Fern\'andez-Prieto}}, \bibinfo {author} {\bibfnamefont {A.}~\bibnamefont
			{Lujambio}}, \bibinfo {author} {\bibfnamefont {V.}~\bibnamefont {Losada}},
		\bibinfo {author} {\bibfnamefont {M.~J.}\ \bibnamefont {Freire}},\ and\
		\bibinfo {author} {\bibfnamefont {A.}~\bibnamefont {Cabello}},\ }\bibfield
	{title} {\bibinfo {title} {Classical physics and the bounds of quantum
			correlations},\ }\href {https://doi.org/10.1103/PhysRevLett.116.250404}
	{\bibfield  {journal} {\bibinfo  {journal} {Phys. Rev. Lett.}\ }\textbf
		{\bibinfo {volume} {116}},\ \bibinfo {pages} {250404} (\bibinfo {year}
		{2016})}\BibitemShut {NoStop}%
	\bibitem [{\citenamefont {Zhang}\ \emph {et~al.}(2019)\citenamefont {Zhang},
		\citenamefont {Xu}, \citenamefont {Xie}, \citenamefont {Zhang}, \citenamefont
		{Smith}, \citenamefont {Kim},\ and\ \citenamefont
		{Zhang}}]{zhang2019experimental}%
	\BibitemOpen
	\bibfield  {author} {\bibinfo {author} {\bibfnamefont {A.}~\bibnamefont
			{Zhang}}, \bibinfo {author} {\bibfnamefont {H.}~\bibnamefont {Xu}}, \bibinfo
		{author} {\bibfnamefont {J.}~\bibnamefont {Xie}}, \bibinfo {author}
		{\bibfnamefont {H.}~\bibnamefont {Zhang}}, \bibinfo {author} {\bibfnamefont
			{B.~J.}\ \bibnamefont {Smith}}, \bibinfo {author} {\bibfnamefont {M.~S.}\
			\bibnamefont {Kim}},\ and\ \bibinfo {author} {\bibfnamefont {L.}~\bibnamefont
			{Zhang}},\ }\bibfield  {title} {\bibinfo {title} {Experimental test of
			contextuality in quantum and classical systems},\ }\href
	{https://doi.org/10.1103/PhysRevLett.122.080401} {\bibfield  {journal}
		{\bibinfo  {journal} {Phys. Rev. Lett.}\ }\textbf {\bibinfo {volume} {122}},\
		\bibinfo {pages} {080401} (\bibinfo {year} {2019})}\BibitemShut {NoStop}%
	\bibitem [{\citenamefont {Fine}(1982)}]{FinePRL1982}%
	\BibitemOpen
	\bibfield  {author} {\bibinfo {author} {\bibfnamefont {A.}~\bibnamefont
			{Fine}},\ }\bibfield  {title} {\bibinfo {title} {Hidden variables, joint
			probability, and the bell inequalities},\ }\href
	{https://doi.org/10.1103/PhysRevLett.48.291} {\bibfield  {journal} {\bibinfo
			{journal} {Phys. Rev. Lett.}\ }\textbf {\bibinfo {volume} {48}},\ \bibinfo
		{pages} {291} (\bibinfo {year} {1982})}\BibitemShut {NoStop}%
	\bibitem [{\citenamefont {Cabello}(2016)}]{cabello2016simple}%
	\BibitemOpen
	\bibfield  {author} {\bibinfo {author} {\bibfnamefont {A.}~\bibnamefont
			{Cabello}},\ }\bibfield  {title} {\bibinfo {title} {Simple method for
			experimentally testing any form of quantum contextuality},\ }\href
	{https://doi.org/10.1103/PhysRevA.93.032102} {\bibfield  {journal} {\bibinfo
			{journal} {Phys. Rev. A}\ }\textbf {\bibinfo {volume} {93}},\ \bibinfo
		{pages} {032102} (\bibinfo {year} {2016})}\BibitemShut {NoStop}%
	\bibitem [{\citenamefont {Kok}\ \emph {et~al.}(2002)\citenamefont {Kok},
		\citenamefont {Lee},\ and\ \citenamefont {Dowling}}]{kok2002single}%
	\BibitemOpen
	\bibfield  {author} {\bibinfo {author} {\bibfnamefont {P.}~\bibnamefont
			{Kok}}, \bibinfo {author} {\bibfnamefont {H.}~\bibnamefont {Lee}},\ and\
		\bibinfo {author} {\bibfnamefont {J.~P.}\ \bibnamefont {Dowling}},\
	}\bibfield  {title} {\bibinfo {title} {Single-photon quantum-nondemolition
			detectors constructed with linear optics and projective measurements},\
	}\href {https://doi.org/10.1103/PhysRevA.66.063814} {\bibfield  {journal}
		{\bibinfo  {journal} {Phys. Rev. A}\ }\textbf {\bibinfo {volume} {66}},\
		\bibinfo {pages} {063814} (\bibinfo {year} {2002})}\BibitemShut {NoStop}%
	\bibitem [{\citenamefont {Imoto}\ \emph {et~al.}(1985)\citenamefont {Imoto},
		\citenamefont {Haus},\ and\ \citenamefont {Yamamoto}}]{imoto1985quantum}%
	\BibitemOpen
	\bibfield  {author} {\bibinfo {author} {\bibfnamefont {N.}~\bibnamefont
			{Imoto}}, \bibinfo {author} {\bibfnamefont {H.~A.}\ \bibnamefont {Haus}},\
		and\ \bibinfo {author} {\bibfnamefont {Y.}~\bibnamefont {Yamamoto}},\
	}\bibfield  {title} {\bibinfo {title} {Quantum nondemolition measurement of
			the photon number via the optical kerr effect},\ }\href
	{https://doi.org/10.1103/PhysRevA.32.2287} {\bibfield  {journal} {\bibinfo
			{journal} {Phys. Rev. A}\ }\textbf {\bibinfo {volume} {32}},\ \bibinfo
		{pages} {2287} (\bibinfo {year} {1985})}\BibitemShut {NoStop}%
	\bibitem [{\citenamefont {Sanders}\ and\ \citenamefont
		{Rice}(1999)}]{sanders1999nonclassical}%
	\BibitemOpen
	\bibfield  {author} {\bibinfo {author} {\bibfnamefont {B.~C.}\ \bibnamefont
			{Sanders}}\ and\ \bibinfo {author} {\bibfnamefont {D.~A.}\ \bibnamefont
			{Rice}},\ }\bibfield  {title} {\bibinfo {title} {Nonclassical fields and the
			nonlinear interferometer},\ }\href
	{https://doi.org/10.1103/PhysRevA.61.013805} {\bibfield  {journal} {\bibinfo
			{journal} {Phys. Rev. A}\ }\textbf {\bibinfo {volume} {61}},\ \bibinfo
		{pages} {013805} (\bibinfo {year} {1999})}\BibitemShut {NoStop}%
	\bibitem [{\citenamefont {Ou}(1996)}]{ou1996complementarity}%
	\BibitemOpen
	\bibfield  {author} {\bibinfo {author} {\bibfnamefont {Z.~Y.}\ \bibnamefont
			{Ou}},\ }\bibfield  {title} {\bibinfo {title} {Complementarity and
			fundamental limit in precision phase measurement},\ }\href
	{https://doi.org/10.1103/PhysRevLett.77.2352} {\bibfield  {journal} {\bibinfo
			{journal} {Phys. Rev. Lett.}\ }\textbf {\bibinfo {volume} {77}},\ \bibinfo
		{pages} {2352} (\bibinfo {year} {1996})}\BibitemShut {NoStop}%
	\bibitem [{\citenamefont {Boyd}(1999)}]{boyd1999order}%
	\BibitemOpen
	\bibfield  {author} {\bibinfo {author} {\bibfnamefont {R.~W.}\ \bibnamefont
			{Boyd}},\ }\bibfield  {title} {\bibinfo {title} {Order-of-magnitude estimates
			of the nonlinear optical susceptibility},\ }\href
	{https://doi.org/10.1080/09500349908231277} {\bibfield  {journal} {\bibinfo
			{journal} {Journal of Modern Optics}\ }\textbf {\bibinfo {volume} {46}},\
		\bibinfo {pages} {367} (\bibinfo {year} {1999})}\BibitemShut {NoStop}%
	\bibitem [{\citenamefont {Lukin}\ and\ \citenamefont
		{Imamo\ifmmode~\breve{g}\else \u{g}\fi{}lu}(2000)}]{lukin2000nonlinear}%
	\BibitemOpen
	\bibfield  {author} {\bibinfo {author} {\bibfnamefont {M.~D.}\ \bibnamefont
			{Lukin}}\ and\ \bibinfo {author} {\bibfnamefont {A.}~\bibnamefont
			{Imamo\ifmmode~\breve{g}\else \u{g}\fi{}lu}},\ }\bibfield  {title} {\bibinfo
		{title} {Nonlinear optics and quantum entanglement of ultraslow single
			photons},\ }\href {https://doi.org/10.1103/PhysRevLett.84.1419} {\bibfield
		{journal} {\bibinfo  {journal} {Phys. Rev. Lett.}\ }\textbf {\bibinfo
			{volume} {84}},\ \bibinfo {pages} {1419} (\bibinfo {year}
		{2000})}\BibitemShut {NoStop}%
	\bibitem [{\citenamefont {Glauber}(1963)}]{glauber1963photon}%
	\BibitemOpen
	\bibfield  {author} {\bibinfo {author} {\bibfnamefont {R.~J.}\ \bibnamefont
			{Glauber}},\ }\bibfield  {title} {\bibinfo {title} {Photon correlations},\
	}\href {https://doi.org/10.1103/PhysRevLett.10.84} {\bibfield  {journal}
		{\bibinfo  {journal} {Phys. Rev. Lett.}\ }\textbf {\bibinfo {volume} {10}},\
		\bibinfo {pages} {84} (\bibinfo {year} {1963})}\BibitemShut {NoStop}%
	\bibitem [{\citenamefont {Sudarshan}(1963)}]{sudarshan1963equivalence}%
	\BibitemOpen
	\bibfield  {author} {\bibinfo {author} {\bibfnamefont {E.~C.~G.}\
			\bibnamefont {Sudarshan}},\ }\bibfield  {title} {\bibinfo {title}
		{Equivalence of semiclassical and quantum mechanical descriptions of
			statistical light beams},\ }\href
	{https://doi.org/10.1103/PhysRevLett.10.277} {\bibfield  {journal} {\bibinfo
			{journal} {Phys. Rev. Lett.}\ }\textbf {\bibinfo {volume} {10}},\ \bibinfo
		{pages} {277} (\bibinfo {year} {1963})}\BibitemShut {NoStop}%
	\bibitem [{\citenamefont {Rahimi-Keshari}\ \emph {et~al.}(2011)\citenamefont
		{Rahimi-Keshari}, \citenamefont {Scherer}, \citenamefont {Mann},
		\citenamefont {Rezakhani}, \citenamefont {Lvovsky},\ and\ \citenamefont
		{Sanders}}]{rahimi2011quantum}%
	\BibitemOpen
	\bibfield  {author} {\bibinfo {author} {\bibfnamefont {S.}~\bibnamefont
			{Rahimi-Keshari}}, \bibinfo {author} {\bibfnamefont {A.}~\bibnamefont
			{Scherer}}, \bibinfo {author} {\bibfnamefont {A.}~\bibnamefont {Mann}},
		\bibinfo {author} {\bibfnamefont {A.~T.}\ \bibnamefont {Rezakhani}}, \bibinfo
		{author} {\bibfnamefont {A.~I.}\ \bibnamefont {Lvovsky}},\ and\ \bibinfo
		{author} {\bibfnamefont {B.~C.}\ \bibnamefont {Sanders}},\ }\bibfield
	{title} {\bibinfo {title} {Quantum process tomography with coherent states},\
	}\href {https://doi.org/10.1088/1367-2630/13/1/013006} {\bibfield  {journal}
		{\bibinfo  {journal} {New Journal of Physics}\ }\textbf {\bibinfo {volume}
			{13}},\ \bibinfo {pages} {013006} (\bibinfo {year} {2011})}\BibitemShut
	{NoStop}%
	\bibitem [{\citenamefont {Reid}\ and\ \citenamefont
		{Walls}(1986)}]{reid1986violations}%
	\BibitemOpen
	\bibfield  {author} {\bibinfo {author} {\bibfnamefont {M.~D.}\ \bibnamefont
			{Reid}}\ and\ \bibinfo {author} {\bibfnamefont {D.~F.}\ \bibnamefont
			{Walls}},\ }\bibfield  {title} {\bibinfo {title} {Violations of classical
			inequalities in quantum optics},\ }\href
	{https://doi.org/10.1103/PhysRevA.34.1260} {\bibfield  {journal} {\bibinfo
			{journal} {Phys. Rev. A}\ }\textbf {\bibinfo {volume} {34}},\ \bibinfo
		{pages} {1260} (\bibinfo {year} {1986})}\BibitemShut {NoStop}%
	\bibitem [{\citenamefont {Catani}\ \emph {et~al.}(2023)\citenamefont {Catani},
		\citenamefont {Leifer}, \citenamefont {Scala}, \citenamefont {Schmid},\ and\
		\citenamefont {Spekkens}}]{catani}%
	\BibitemOpen
	\bibfield  {author} {\bibinfo {author} {\bibfnamefont {L.}~\bibnamefont
			{Catani}}, \bibinfo {author} {\bibfnamefont {M.}~\bibnamefont {Leifer}},
		\bibinfo {author} {\bibfnamefont {G.}~\bibnamefont {Scala}}, \bibinfo
		{author} {\bibfnamefont {D.}~\bibnamefont {Schmid}},\ and\ \bibinfo {author}
		{\bibfnamefont {R.~W.}\ \bibnamefont {Spekkens}},\ }\bibfield  {title}
	{\bibinfo {title} {Aspects of the phenomenology of interference that are
			genuinely nonclassical},\ }\href
	{https://doi.org/10.1103/PhysRevA.108.022207} {\bibfield  {journal} {\bibinfo
			{journal} {Phys. Rev. A}\ }\textbf {\bibinfo {volume} {108}},\ \bibinfo
		{pages} {022207} (\bibinfo {year} {2023})}\BibitemShut {NoStop}%
	\bibitem [{\citenamefont {Wagner}\ \emph {et~al.}(2024)\citenamefont {Wagner},
		\citenamefont {Camillini},\ and\ \citenamefont
		{Galv{\~{a}}o}}]{Wagner2024coherence}%
	\BibitemOpen
	\bibfield  {author} {\bibinfo {author} {\bibfnamefont {R.}~\bibnamefont
			{Wagner}}, \bibinfo {author} {\bibfnamefont {A.}~\bibnamefont {Camillini}},\
		and\ \bibinfo {author} {\bibfnamefont {E.~F.}\ \bibnamefont {Galv{\~{a}}o}},\
	}\bibfield  {title} {\bibinfo {title} {Coherence and contextuality in a
			{M}ach-{Z}ehnder interferometer},\ }\href
	{https://doi.org/10.22331/q-2024-02-05-1240} {\bibfield  {journal} {\bibinfo
			{journal} {{Quantum}}\ }\textbf {\bibinfo {volume} {8}},\ \bibinfo {pages}
		{1240} (\bibinfo {year} {2024})}\BibitemShut {NoStop}%
	\bibitem [{\citenamefont {Rahimi-Keshari}\ \emph {et~al.}(2020)\citenamefont
		{Rahimi-Keshari}, \citenamefont {Baghbanzadeh},\ and\ \citenamefont
		{Caves}}]{rahimi2020}%
	\BibitemOpen
	\bibfield  {author} {\bibinfo {author} {\bibfnamefont {S.}~\bibnamefont
			{Rahimi-Keshari}}, \bibinfo {author} {\bibfnamefont {S.}~\bibnamefont
			{Baghbanzadeh}},\ and\ \bibinfo {author} {\bibfnamefont {C.~M.}\ \bibnamefont
			{Caves}},\ }\bibfield  {title} {\bibinfo {title} {In situ characterization of
			linear-optical networks in randomized boson sampling},\ }\href
	{https://doi.org/10.1103/PhysRevA.101.043809} {\bibfield  {journal} {\bibinfo
			{journal} {Phys. Rev. A}\ }\textbf {\bibinfo {volume} {101}},\ \bibinfo
		{pages} {043809} (\bibinfo {year} {2020})}\BibitemShut {NoStop}%
\end{thebibliography}

%

\end{document}